\documentclass[11pt,a4paper]{article}
\usepackage{amsmath}
\usepackage{soul}

\usepackage{booktabs} 
\usepackage{enumitem} 
\usepackage{mathrsfs}
\usepackage{rotating}
\usepackage[natbibapa]{apacite}
\defcitealias{camsdata}{INERIS et.~al, 2022}
\defcitealias{nasadataset}{CIESIN et.~al, 2018}
\defcitealias{meteoseasons}{NCEI, 2023}
\defcitealias{cds}{CDS, 2020}
\defcitealias{cds2}{CDS, 2021} 

\usepackage[pdftex,colorlinks,citecolor=blue,urlcolor=blue,linkcolor=red]{hyperref} 

\usepackage{doi}

\usepackage{graphicx} 
\usepackage[small,bf,hang]{caption}	
\usepackage{subcaption}
\usepackage{nicefrac}
\usepackage{tabularx}
\usepackage[export]{adjustbox}
\usepackage{afterpage}

\usepackage{amssymb} 
\usepackage{amsmath} 
\usepackage{accents}
\usepackage{bbm} 
\usepackage{amsbsy} 
\usepackage{array}
\usepackage{multirow} 
\usepackage[dvipsnames,table]{xcolor} 
\usepackage{mathtools}
\usepackage{eurosym}

\usepackage{floatrow}
\newfloatcommand{capbtabbox}{table}[][\FBwidth]

\usepackage[ruled]{algorithm2e} 
\SetAlCapFnt{\small} 
\SetKw{KwTo}{in}

\usepackage{parskip}
\usepackage{authblk} 
\usepackage{footnote}

\bibpunct[, ]{(}{)}{;}{a}{,}{,}

\usepackage{tikz,pgfplots}
\usetikzlibrary{fit, positioning, chains,shapes.arrows,fit}
\usetikzlibrary{shapes, arrows, arrows.meta}
\usetikzlibrary{decorations.pathreplacing}
\usetikzlibrary{calc} 
\usetikzlibrary{matrix} 
\usepackage{pgf}
\usetikzlibrary{decorations.text}
\tikzset{cross/.style={cross out, draw=black, minimum size=2*(#1-\pgflinewidth), inner sep=0pt, outer sep=0pt},
cross/.default={1pt}}
\usetikzlibrary{shapes.misc}
\usepackage{environ}
\makeatletter
\newsavebox{\measure@tikzpicture}
\NewEnviron{scaletikzpicturetowidth}[1]{%
  \def\tikz@width{#1}%
  \begin{lrbox}{\measure@tikzpicture}%
  \BODY
  \end{lrbox}%
  \pgfmathparse{#1/\wd\measure@tikzpicture}%
  \BODY
}
\makeatother
\tikzdeclarecoordinatesystem{page}{
    \parsecomma#1\endparsecomma
    \pgfpointanchor{current page}{north east}
    \pgf@xc=\pgf@x%
    \pgf@yc=\pgf@y%
    \pgfpointanchor{current page}{south west}
    \pgf@xb=\pgf@x%
    \pgf@yb=\pgf@y%
    \pgfmathparse{(\pgf@xc-\pgf@xb)/2.*\page@x+(\pgf@xc+\pgf@xb)/2.}
    \expandafter\pgf@x\expandafter=\pgfmathresult pt
    \pgfmathparse{(\pgf@yc-\pgf@yb)/2.*\page@y+(\pgf@yc+\pgf@yb)/2.}
    \expandafter\pgf@y\expandafter=\pgfmathresult pt
}
\makeatother
\usetikzlibrary{arrows,automata}
\usepackage{ragged2e}

\tikzset{
    max width/.style args={#1}{
        execute at begin node={\begin{varwidth}{#1}},
        execute at end node={\end{varwidth}}
    }
}
\definecolor{newblue}{RGB}{86, 180, 233}
\definecolor{newred}{RGB}{248, 118, 109}
\definecolor{newgreen}{RGB}{163, 213, 150}

\colorlet{textcolor}{white}
\colorlet{bordercolor}{white}

\pgfdeclarelayer{background}
\pgfsetlayers{background,main}

\definecolor{airforceblue}{rgb}{0.36, 0.54, 0.66}
\definecolor{forestgreen}{rgb}{0.13, 0.55, 0.13}\definecolor{fulvous}{rgb}{0.86, 0.52, 0.0}
\definecolor{gray}{rgb}{0.5, 0.5, 0.5}
\definecolor{bistre}{rgb}{0.24, 0.17, 0.12}\definecolor{bostonuniversityred}{rgb}{0.8, 0.0, 0.0}
\definecolor{purpleheart}{rgb}{0.41, 0.21, 0.61}
\definecolor{lightsalmonpink}{rgb}{1.0, 0.6, 0.6}\definecolor{arrowcolor}{rgb}{0.92, 0.92, 0.92}
\tikzset{
inner/.style={
  on chain,
  circle,
  inner sep=4pt,
  fill=circlecolor,
  line width=1.5pt,
  draw=bordercolor,
  text width=1.2em,
  align=center,
  text height=1.25ex,
  text depth=0ex
},
on grid
}
\usepackage{wallpaper}
\usepackage{pdfpages}

\usepackage{regexpatch}
\usepackage[final,defaultcolor=orange]{changes}

\usepackage{pbox}
\newcommand\drawarrow{
\node[on chain] (f) {};
\begin{pgfonlayer}{background}
\node[
  inner sep=10pt,
  single arrow,
  single arrow head extend=0.6cm,
  draw=none,
  fill=arrowcolor,
  fit= (c1) (f)
] (arrow) {};
\fill[white] 
  (arrow.before tail) -- (c1|-arrow.west) -- (arrow.after tail) -- cycle;
\end{pgfonlayer}
}

\numberwithin{equation}{section}

\usepackage{relsize}

\usepackage{setspace}
\def\spacingset#1{\renewcommand{\baselinestretch}%
{#1}\small\normalsize} \spacingset{1}

\makeatletter
\let\ref\@refstar
\makeatother

\begin{document}
\def\spacingset#1{\renewcommand{\baselinestretch}%
{#1}\small\normalsize} \spacingset{1}

\title{\bf{Socio-demographic inequalities in the maximum human lifespan: evidence from Belgium and the Netherlands}}

\author[1,*,$\dagger$]{Jens Robben}
\author[1,$\dagger$]{Torsten Kleinow}
\affil[1]{Research Centre for Longevity Risk, Faculty of Economics and Business, University of Amsterdam, 1012 WP Amsterdam, The Netherlands.}
\affil[$\dagger$]{These authors contributed equally to this work.}
\affil[*]{Corresponding author: \href{mailto:j.robben@uva.nl}{j.robben@uva.nl}}
\maketitle
\thispagestyle{empty}

\begin{abstract} \noindent
The existence of an upper limit to the human lifespan has been widely debated, with studies offering both supporting and opposing evidence. Existing studies typically treat this limit, if it exists, as a population-wide constant, overlooking potential heterogeneity across individuals. Using unique individual-level death and population records for all individuals aged 90 and older in Belgium and the Netherlands between 1995 and 2022, \added{we use extreme value theory to estimate the upper endpoint of the lifespan distribution, conditional on survival beyond a high threshold age. We find that this endpoint is finite in both countries, but that it is not well described by a single population-wide value: differences in sex, household type, civil status, and origin translate into differences of several years in the estimated upper endpoint. We conclude that socio-demographic inequality persists even at the most advanced ages.}
\end{abstract}

\noindent%

\section{Introduction}
Life expectancy has increased steadily in most high-income countries over the past century \citep{wilmoth2000demography, bongaarts2006long}. Historically, these gains came from reductions in early-life mortality, but since the mid-twentieth century the main driver has been declining mortality at older ages \citep{rau2008continued, vaupel1997remarkable}. These trends have increased interest in the question of whether these continued life expectancy gains reflect a postponement toward a fixed upper limit in the human lifespan, or a shift in the upper limit itself. Empirical evidence remains mixed, with studies reaching conflicting conclusions \citep{lenart2017questionable, dong2016evidence, zuo2018advancing, weon2009theoretical}.

Much of the debate has treated the lifespan limit (if any) as a population-wide constant. This assumption contrasts with well-established evidence of socio-demographic disparities in mortality throughout the life course \citep{mackenbach1997socioeconomic, wen2023modelling, balia2008mortality, hobcraft1984socio, cairns2019modelling, nigri2026measuring}. While such differences are especially large at younger and adult ages, it is unknown whether they persist among the oldest-old ages, where mortality selection is strongest and data are scarce. In particular, little is known about whether the upper tail of the lifespan distribution varies systematically across population subgroups. While some studies allow for individual heterogeneity in the lifespan limit \citep{huang2021mixture}, whether it varies across socio-demographic groups remains an open question. This question is timely since the number of centenarians is expected to exceed 25 million by 2100, and understanding the social determinants of extreme longevity provides insights into health inequalities \citep{robine2017worldwide}.

In this paper, we study the upper tail of the lifespan distribution using population-wide individual-level data from Belgium and the Netherlands, covering all individuals aged 90 and older between 1995 and 2022. These microdata include information on sex, origin, civil status, household type, and education, allowing us to study the maximum lifespan in greater depth than previous studies. 

Several studies find evidence for a finite lifespan limit \citep{aarssen1994maximal, hanayama2016estimating, gbari2017extreme, einmahl2019limits, huang2020modelling, barbi2018plateau}, while others argue against the existence of such a limit \citep{watts2006extreme, rootzen2017human} or that the evidence remains inconclusive \citep{ferreira2018human, gavrilova2020we}. Our results provide evidence consistent with a finite upper endpoint of the lifespan distribution. More importantly, we show that this endpoint is not well described by a single value: it varies substantially across socio-demographic groups. 

These findings imply that the concept of a universal lifespan limit may be misleading. Even among individuals who survive to very old ages,  socio-demographic disparities are associated with substantial variation in the upper limit of the maximum lifespans. Moving beyond a single-number characterization provides new insights into the social stratification of extreme longevity.

\section{Materials and methods} \label{sec:data}

We use population-wide individual-level microdata covering all individuals aged 90 and older who resided in Belgium or the Netherlands at any point between 1995 and 2022. The data combine complete population registers and death records with socio-demographic information, including sex, origin, civil status, household type, and educational attainment (Table~\ref{tab:socdemcov}). 

\begin{table}[!ht]
\centering
\begin{tabularx}{\textwidth}{l X}
\toprule
Covariate & Description \\
\midrule

\texttt{sex} & Sex: \texttt{male} or \texttt{female} \\

\addlinespace
\texttt{org} & Region of origin: 
\texttt{native} (Belgian/Dutch); 
\texttt{west-europe} (Western Europe); 
\texttt{non-western} (outside Western Europe) \\

\addlinespace
\texttt{civ} & Civil or marital status: 
\texttt{widowed} or \texttt{divorced} (post-marriage/partnership); 
\texttt{single} (never married/no partnership); 
\texttt{married} (married/partnership) \\

\addlinespace
\texttt{hht} & Household type: 
\texttt{collective} (collective/institutional); 
\texttt{single-person}; 
\texttt{couple} (couple without children); 
\texttt{family} (couple or single-parent with children); 
\texttt{other} (other private household composition) \\

\addlinespace
\texttt{edu} & Highest education completed (or of children): 
\texttt{unobserved} (missing observations, including all pre-2001 observations); 
\texttt{primary} (Belgium ISCED 0--1, Netherlands SOI 11); 
\texttt{secondary} (Belgium ISCED 2--4, Netherlands SOI 12, 21); 
\texttt{tertiary} (Belgium ISCED 5--7, Netherlands SOI 31, 32) \\

\bottomrule
\end{tabularx}
\caption{Socio-demographic covariates \replaced{available in the Belgian and Dutch microdata.}{} \label{tab:socdemcov}}
\end{table}

Educational attainment is harmonized across countries by grouping the International Standard Classification of Education (ISCED) levels in Belgium and the Dutch Standard Classification of Education (SOI) levels in the Netherlands into three categories: primary, secondary, and tertiary education. Because education is unavailable in Belgium for individuals who died before 2001, these observations are treated as missing in both countries for consistency. Missing education at very old ages is common and is imputed for the post-2001 period using the highest educational attainment of the individual's children when available.

Individuals are followed from age 90 or from their first entry into the dataset if this occurs later. Follow-up continues until death or right censoring due to emigration or survival beyond 2022. The data are subject to left truncation, as individuals from older cohorts enter observation only if they survive to the start of the study period. These features are explicitly accounted for in the statistical analysis (Appendix~\ref{app:likelihood}). \added{As in any empirical study we rely on the accuracy of the recorded dates. Dates of birth and death are taken from the administrative population registers, and we discuss their registration and validation, as well as the treatment of partially observed dates of birth, in Appendix~\ref{app:validation}.}

\subsection{Belgian microdata}

The Belgian microdata provided by Statbel, the Belgian statistical office, include $629\,480$ individuals aged 90+, of whom 78.8\% died during the observation period and 21.2\% are right-censored. Censoring occurs almost entirely in 2022, leading to a concentration of observations in the final period. The number of individuals reaching very advanced ages increases substantially over time (Table~\ref{tab:sumbel}).

\begin{table}[!ht]
\centering
\adjustbox{max width=\textwidth}{%
\begin{tabular}[t]{lrrrrrrr|r}
\toprule
Age/Year & 1995–1998 & 1999–2002 & 2003–2006 & 2007–2010 & 2011–2014 & 2015–2018 & 2019–2022 & \textbf{Overall}\\
\midrule
106+      & 26 & 51 & 85 & 65 & 106 & 126 & 189 & \textbf{648}\\
(104,106] & 92  & 137 & 186 & 244 & 298 & 351 & 426 & \textbf{1\ 734}\\
(102,104] & 375 & 489 & 577 & 663 & 847 & 961 & 1\ 429 & \textbf{5\ 341} \\
(100,102] & 989 & 1\ 254 & 1\ 583 & 1\ 831 & 2\ 119 & 1\ 955 & 4\ 716 & \textbf{14\ 447}\\
(98,100] & 2\ 174 & 2\ 966 & 3\ 464 & 3\ 680 & 4\ 321 & 3\ 848 & 11\ 842 & \textbf{32\ 295} \\
(96,98] & 4\ 707 & 5\ 856 & 6\ 726 & 7\ 311 & 6\ 684 & 9\ 419 & 23\ 634 & \textbf{64\ 337}\\
(94,96] & 8\ 665 & 10\ 555 & 11\ 322 & 12\ 083 & 10\ 013 & 18\ 260 & 40\ 564 & \textbf{111\ 462}\\
(92, 94] & 13\ 826 & 15\ 831 & 17\ 098 & 14\ 354 & 18\ 927 & 25\ 425 & 63\ 671 & \textbf{169\ 132}\\
(90,92] & 19\ 431 & 21\ 825 & 22\ 100 & 16\ 829 & 28\ 709 & 31\ 116 & 90\ 074 & \textbf{230\ 084}\\ \midrule
\textbf{Overall} & \textbf{50\ 285} & \textbf{58\ 964} & \textbf{63\ 141} & \textbf{57\ 060} & \textbf{72\ 024} & \textbf{91\ 461} & \textbf{236\ 545} & \textbf{629\ 480}\\
\bottomrule
\end{tabular}}
\caption{Number of individuals by age at last observation (death or right censoring) and calendar period in Belgium. Source: Statbel and authors' calculations. \label{tab:sumbel}}
\end{table}

\begin{figure}[!ht]
    \centering
    \includegraphics[width=0.9\linewidth]{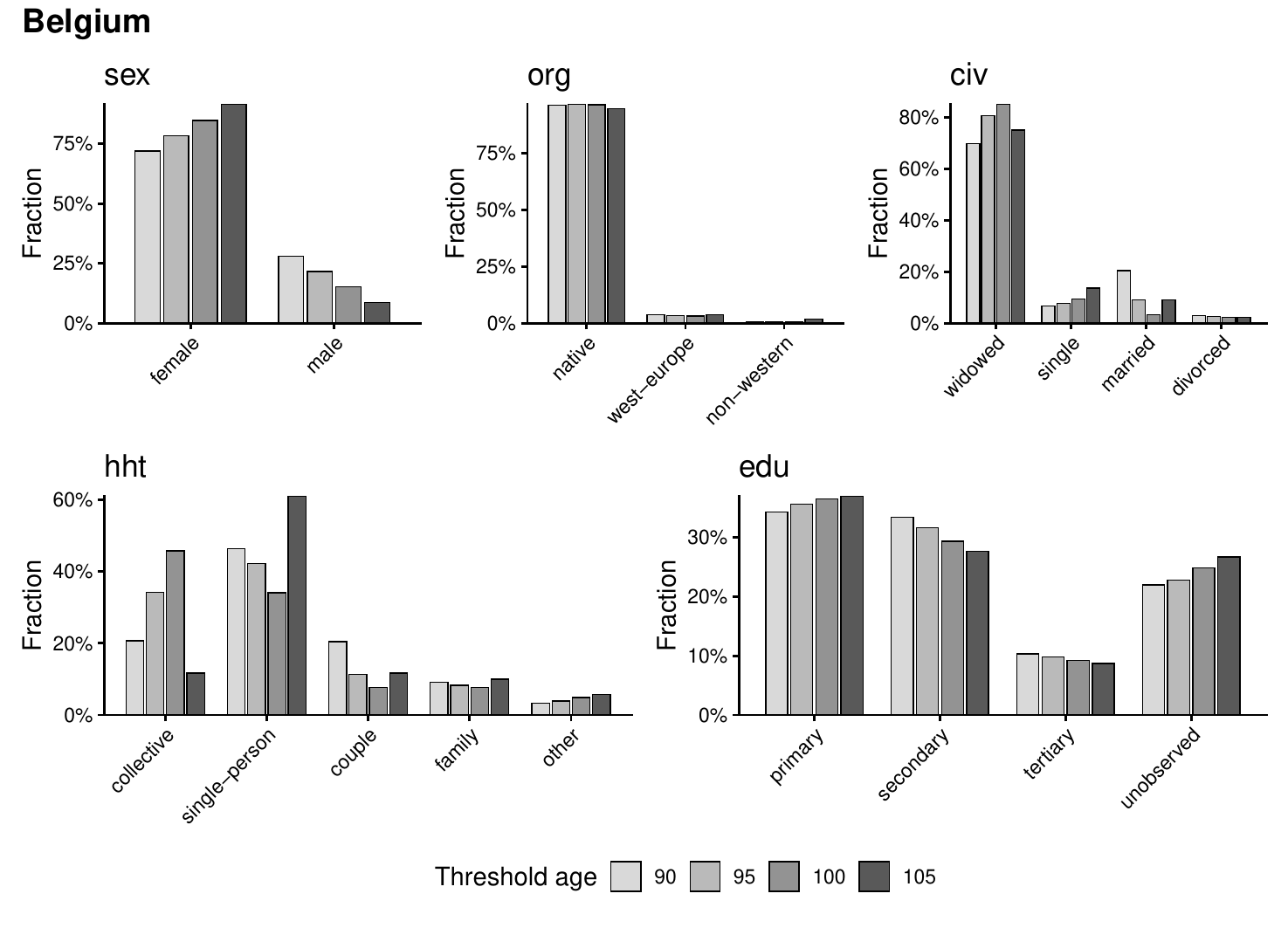}
    \caption{Barplots of the socio-demographic covariates at different threshold ages in Belgium. Source: Statbel and authors' calculations.}
    \label{fig:cov.belgium}
\end{figure}

Figure~\ref{fig:cov.belgium} shows the distribution of socio-demographic characteristics at ages 90, 95, 100, and 105. The proportion of females increases strongly with age, rising from about 72\% at age 90 to more than 90\% at age 105. Widowhood becomes increasingly dominant at older ages, while the share of married individuals declines. Residence in collective households increases markedly with age, consistent with institutional care at the oldest ages. Educational distributions remain relatively stable across thresholds.

\subsection{Dutch microdata}

The Dutch microdata provided by Statistics Netherlands (CBS) comprise $776\,595$ individuals aged 90+, of whom 82.6\% died during follow-up and 17.4\% are right-censored. As in Belgium, censoring occurs almost entirely in the final observation year, and the number of very old individuals increases strongly over time (Table~\ref{tab:sumnld}).

\begin{table}[!ht]
\centering
\adjustbox{max width=\textwidth}{%
\begin{tabular}[t]{lrrrrrrr|r}
\toprule
Age/Year & 1995–1998 & 1999–2002 & 2003–2006 & 2007–2010 & 2011–2014 & 2015–2018 & 2019–2022 & \textbf{Overall}\\
\midrule
106+      & 45 & 68 & 63 & 72 & 118 & 131 & 185 & \textbf{682}\\
(104,106] & 170 & 156 & 215 & 239 & 342 & 364 & 539 & \textbf{2\ 025}\\
(102,104] & 523 & 594 & 692 & 765 & 964 & 1\ 155 & 1\ 828 & \textbf{6\ 521} \\
(100,102] & 1\ 476 & 1\ 590 & 1\ 774 & 2\ 098 & 2\ 547 & 2\ 912 & 5\ 356 & \textbf{17\ 753}\\ 
(98,100] & 3\ 208 & 3\ 667 & 4\ 074 & 4\ 667 & 5\ 650 & 6\ 281 & 12\ 838 & \textbf{40\ 385} \\
(96,98] & 6\ 550 & 7\ 486 & 8\ 028 & 8\ 832 & 10\ 227 & 13\ 054 & 26\ 006 & \textbf{80\ 183}\\
(94,96] & 11\ 459 & 13\ 089 & 13\ 678 & 15\ 127 & 16\ 659 & 22\ 514 & 45\ 011 & \textbf{137\ 537}\\
(92, 94] & 17\ 669 & 19\ 790 & 20\ 735 & 21\ 357 & 25\ 623 & 31\ 713 & 71\ 611 & \textbf{208\ 498}\\
(90,92] & 24\ 477 & 26\ 693 & 27\ 709 & 27\ 782 & 35\ 060 & 38\ 475 & 102\ 815 & \textbf{283\ 011}\\ \midrule
\textbf{Overall} & \textbf{65\ 577} & \textbf{73\ 133} & \textbf{76\ 968} & \textbf{80\ 939} & \textbf{97\ 190} & \textbf{116\ 599} & \textbf{266\ 189} & \textbf{776\ 595}\\
\bottomrule
\end{tabular}}
\caption{Number of individuals by age at last observation (death or right censoring) and calendar period in the Netherlands. Source: CBS and authors' calculations. \label{tab:sumnld}}
\end{table}

\begin{figure}[!ht]
    \centering
    \includegraphics[width=0.9\linewidth]{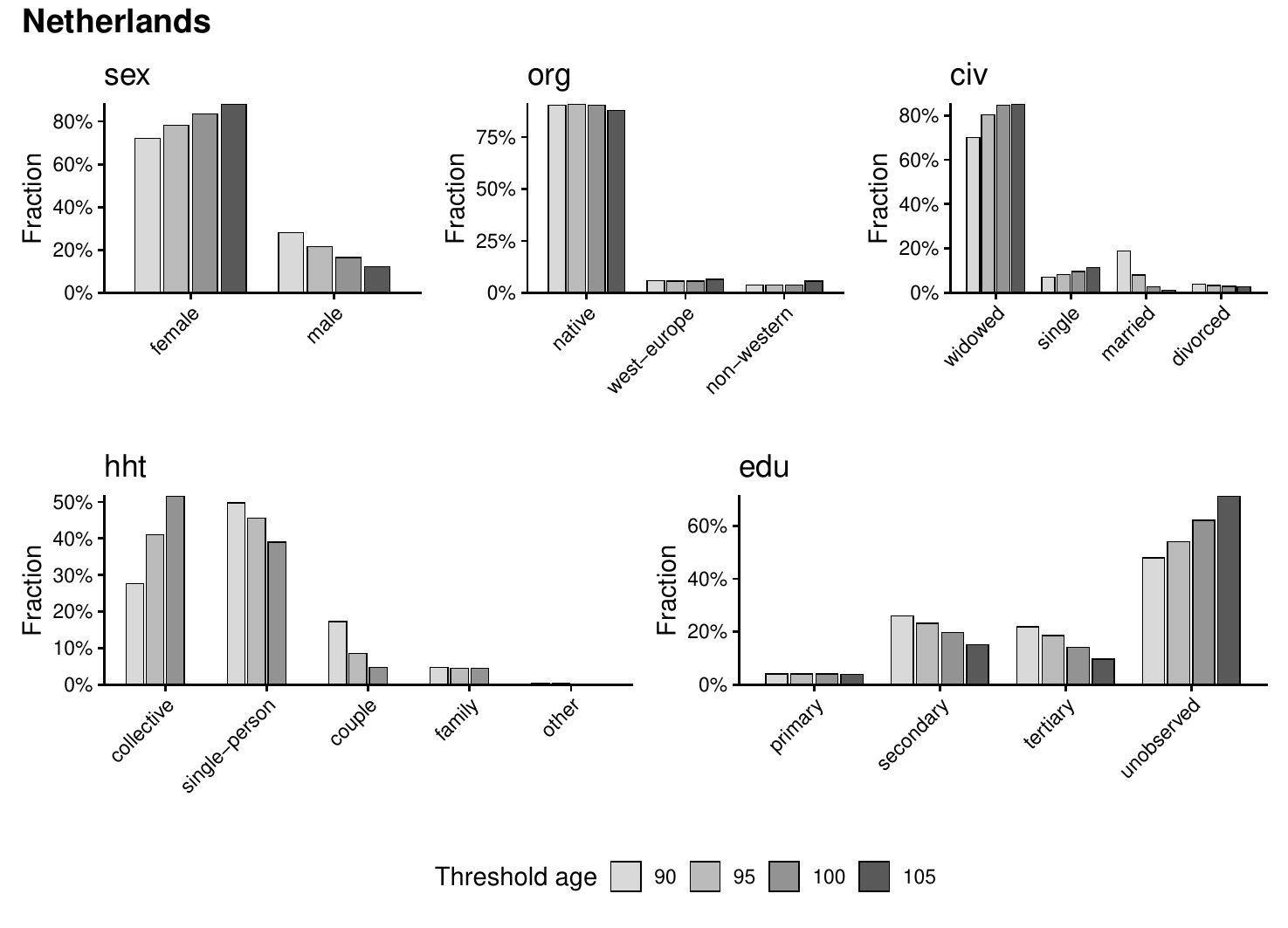}
    \caption{Barplots of the socio-demographic covariates at different threshold ages in the Netherlands. Source: CBS and authors' calculations.}
    \label{fig:cov.Netherlands}
\end{figure}

The evolution of socio-demographic composition is broadly similar to Belgium (Figure~\ref{fig:cov.Netherlands}). The share of females increases with age, and widowhood becomes more prevalent at higher ages. However, important differences emerge: a larger fraction of individuals resides in collective households, and educational attainment is less directly comparable due to the use of children’s education as a proxy for missing parental education. This likely inflates the share of higher education categories in the Dutch data.

\subsection{Ethics statement}
This retrospective study was approved by the Economics and Business Ethics Committee (EBEC) of the University of Amsterdam (approval number EB-10429). The pseudonymized microdata, provided by Statbel (the Belgian statistical office) and Statistics Netherlands (CBS), were accessed for research purposes between 1 April 2025 and 31 August 2026. At no point during or after data collection did the authors have access to information that could directly identify individual participants (e.g., names, addresses, or national identification numbers). All analyses were conducted within secure microdata environments in compliance with the statistical offices' confidentiality regulations.

\subsection{Modeling framework} \label{sec:methods}

We model the upper tail of the lifespan distribution using extreme value theory (EVT). Let $X$ denote age at death. For a sufficiently high threshold $u$, exceedances $Y = X - u$, conditional on $X > u$, are well approximated by a generalized Pareto distribution (GPD).

The tail behavior is governed by a shape parameter $\xi$. When $\xi < 0$, the implied distribution has a finite upper endpoint given by
\begin{equation} \label{eq:maxlifespan}
x^* = u - \frac{\sigma}{\xi}.
\end{equation}
We use this property to assess whether the data support a bounded or unbounded lifespan distribution (Appendix~\ref{app:evt}).

\subsection{Socio-demographic heterogeneity}

To capture heterogeneity in extreme lifespans across population subgroups, we allow the scale parameter of the GPD to depend on socio-demographic characteristics. We model the scale as a log-linear function of covariates \added{selected by forward stepwise selection: starting from an intercept-only scale, we add at each step the covariate that yields the largest reduction in the Akaike Information Criterion (AIC), and stop when no remaining covariate lowers it.} Furthermore, we assume a constant shape parameter, which avoids unrealistic scenarios in which some subgroups would have a statistically implied finite maximum lifespan while others could live indefinitely:
\begin{equation}
\xi_i = \xi, 
\qquad 
\log(\sigma_i) = \beta_0 + \boldsymbol{\beta}_\sigma^\top \mathbf{z}_i,
\end{equation}
where $\mathbf{z}_i$ \added{contains the retained covariates from Table~\ref{tab:socdemcov}, recorded at the age at which an individual enters the risk set of lifetimes exceeding the threshold age $u$.} In other words, this specification allows for heterogeneous dispersion in extreme ages, but assumes that the fundamental tail behavior of the human lifespan distribution is similar across socio-demographic subgroups. \added{This assumption is also supported empirically: allowing $\xi$ to vary across subgroups did not improve model fit, which further justifies the choice for a common $\xi$ (Appendix~\ref{app:sensshape}).}

For each variable, we choose the category with the highest exposure in the Belgian data set as the reference category and use these same categories as references in the Dutch data set. As a result, $\beta_0$ represents the baseline log-scale parameter for an individual in all reference categories.

\subsection{Estimation and censoring adjustment}

The data are subject to right censoring and left truncation due to delayed entry and survival-based inclusion. These features are incorporated directly in the likelihood to ensure valid inference on the upper tail. We provide technical details of the likelihood under truncation and censoring in Appendix~\ref{app:likelihood}.

We estimate model parameters by maximum likelihood, and assess uncertainty using standard likelihood-based methods (Appendices~\ref{app:fullloglik} and~\ref{app:mle}).

\section{Results} \label{sec:results}
\subsection{Assessment of the generalized Pareto approximation} \label{subsec:QQ.GPD}
We first assess whether the generalized Pareto approximation provides an adequate description of the upper tail of the lifespan distribution. We select a threshold age of $u = 100$ years, which is sufficiently high for the asymptotic approximation to be plausible, while still retaining a large number of observations (Belgium: $22\,170$; Netherlands: $26\,945$), particularly when stratifying by socio-demographic characteristics.

\begin{figure}[!ht]
    \centering
    \includegraphics[width=0.45\linewidth]{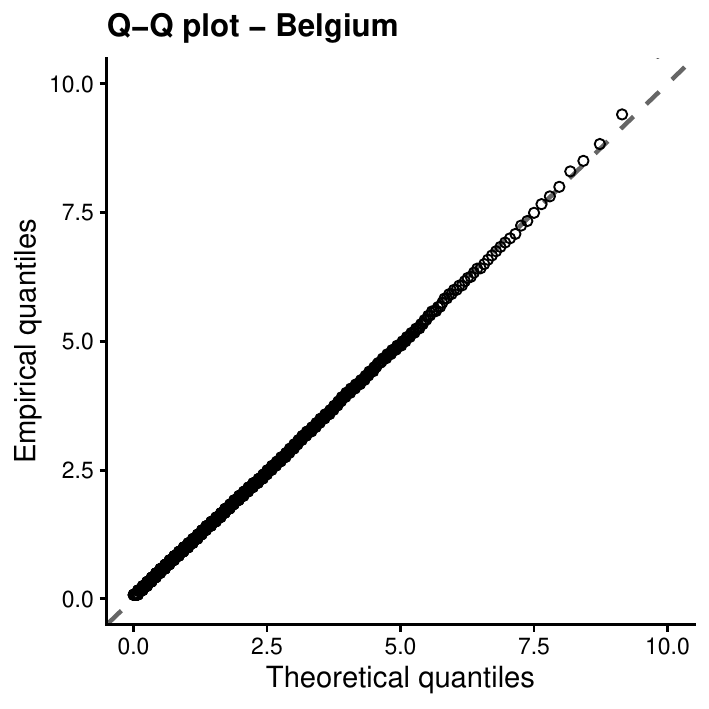}
    \includegraphics[width=0.45\linewidth]{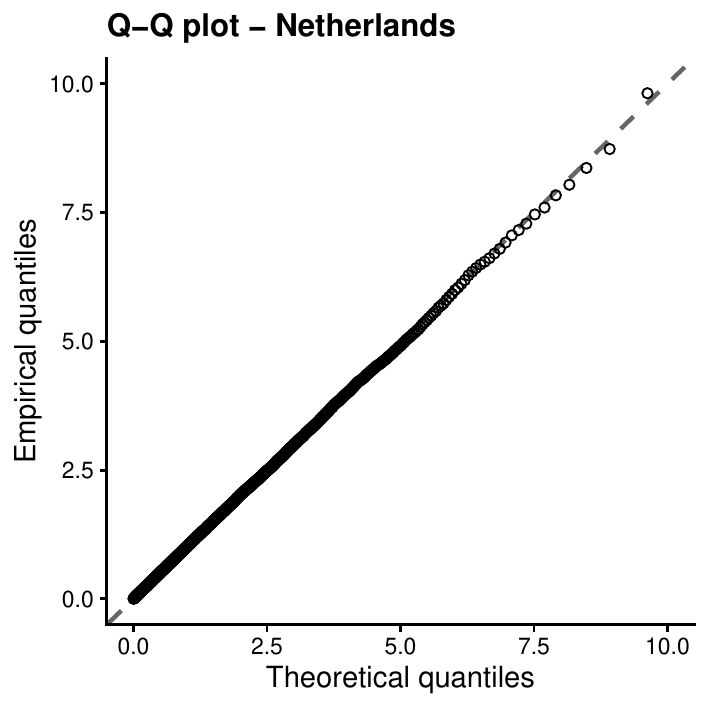}
    \caption{Generalized Pareto Q–Q plots for Belgium (left panel) and the Netherlands (right panel) at a threshold age of 100 years. Source: Statbel, CBS, and authors' calculations.}
    \label{fig:qqplots}
\end{figure}

Figure~\ref{fig:qqplots} presents quantile--quantile (Q--Q) plots comparing theoretical quantiles from the estimated generalized Pareto distribution with the empirical quantiles of the observed exceedances. Across the full range of probabilities, the empirical and theoretical quantiles are closely aligned in both countries. This indicates that the generalized Pareto distribution provides a good approximation to the upper tail at the chosen threshold. Sensitivity analyses with respect to the threshold age confirm that this conclusion is robust (Appendix~\ref{app:sensitivity}). 

\subsection{Covariate selection and goodness of fit}
\added{We determine which socio-demographic characteristics enter the scale parameter, using forward stepwise selection based on the AIC at a threshold age of 100 years. In both Belgium and the Netherlands, household type, sex, civil status, and origin each reduce the AIC and are retained. Educational attainment does not improve the AIC in either country and is therefore excluded from the main specification. A possible reason is that education is unobserved for all pre-2001 observations, and in the Netherlands it is partly derived from the educational attainment of an individual's children. We report a sensitivity analysis including education in Appendix~\ref{app:sensedu}.}

\added{Residual diagnostics for the resulting final covariate model, based on Cox--Snell residuals that account for left truncation and right censoring, confirm an adequate fit both overall and within socio-demographic subgroups (Appendix~\ref{app:diagnostics}).}

\subsection{Evidence for a finite lifespan limit}
We next estimate the extreme value index (EVI) $\xi$ and the covariate effects in the scale parameter by maximizing the log-likelihood. For Belgium, we obtain \added{$\hat{\xi}_{\text{BE}} = -0.1354$ (95\% CI: $[-0.1455, -0.1253]$), and for the Netherlands $\hat{\xi}_{\text{NL}} = -0.1138$ (95\% CI: $[-0.1245, -0.1030]$)}.

In both countries, the EVI is negative, which indicates a light-tailed distribution with a finite upper endpoint. This provides statistical evidence for a bounded human lifespan, consistent across both Belgium and the Netherlands and in line with previous findings in the literature \citep{aarssen1994maximal,hanayama2016estimating, gbari2017extreme, einmahl2019limits}. \added{Adding education to the model or estimating the model on pre-pandemic data (1995--2019) leads to almost identical estimated shape parameters in both countries, see Appendices~\ref{app:sensedu} and~\ref{app:senscovid}, respectively.}

\subsection{Socio-demographic variation in extreme lifespans}

\begin{figure}[!ht]
    \centering
    \includegraphics[width=0.9\linewidth]{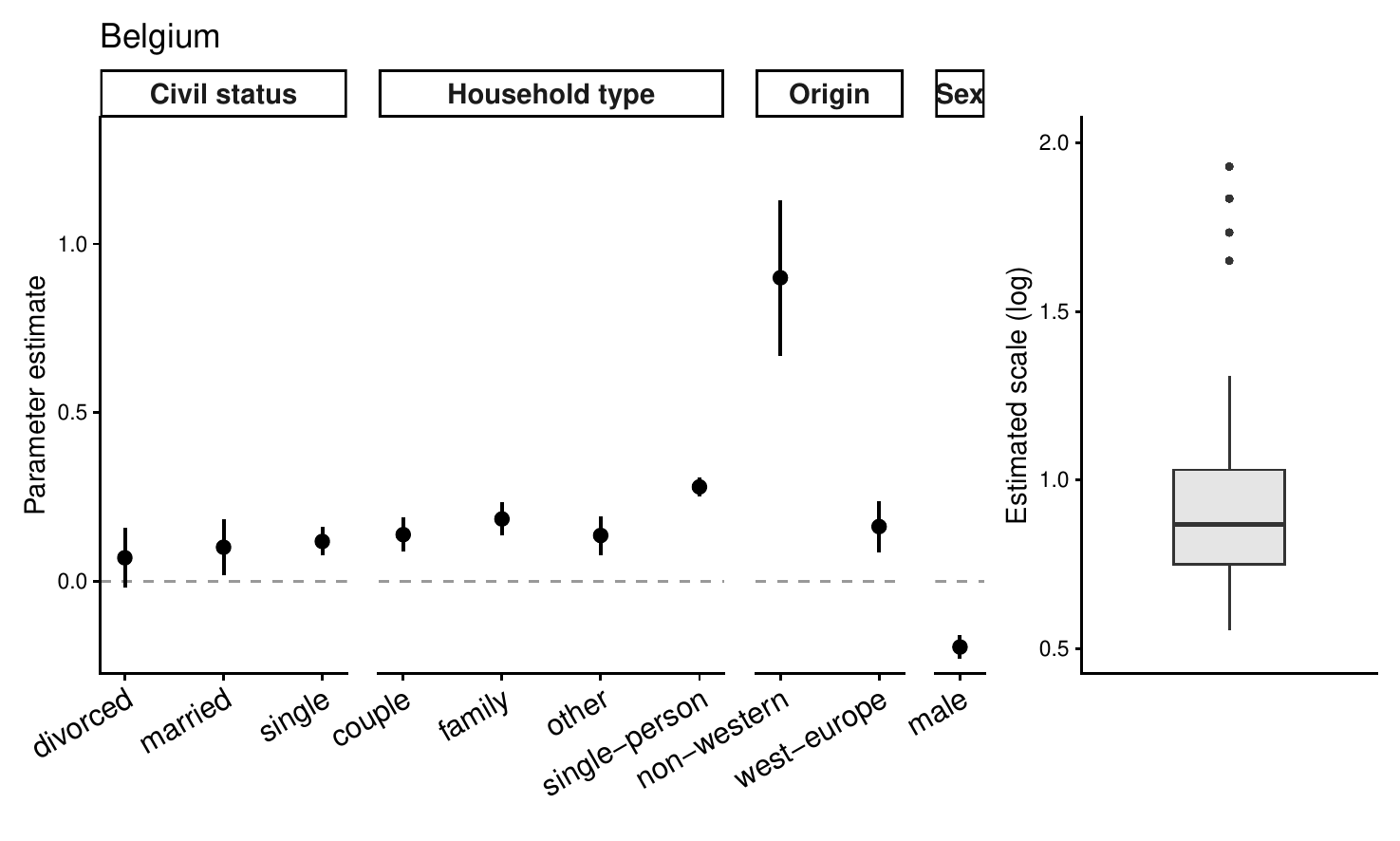}
    \includegraphics[width=0.9\linewidth]{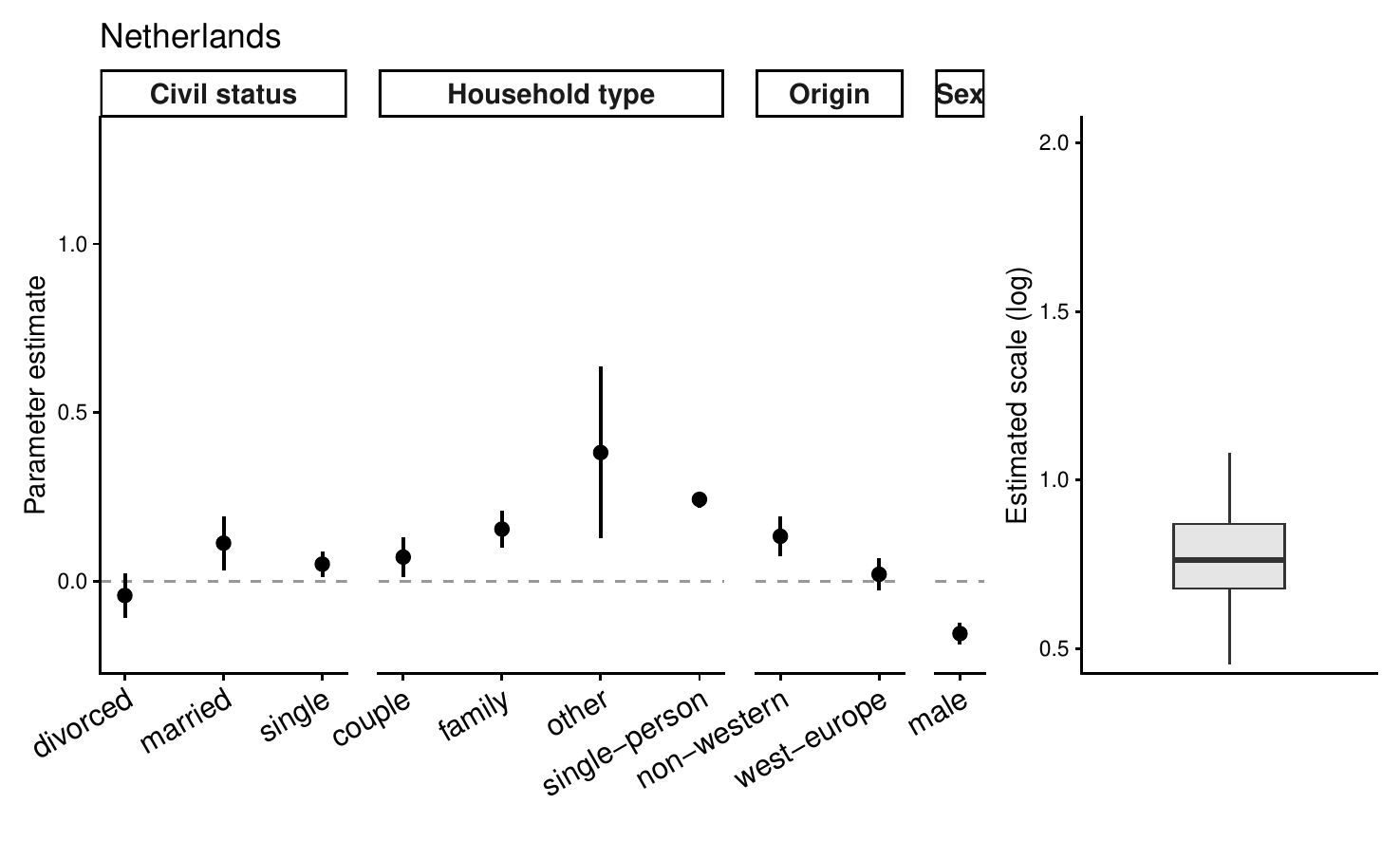}
    \caption{Left panels: Estimated parameters of the socio-demographic covariates in the scale parameter of the generalized Pareto distribution for Belgium (top) and the Netherlands (bottom), with 95$\%$ confidence intervals (Appendix~\ref{app:CIs}, Table~\ref{tabA:cis}). \added{The intercept (0.750 in Belgium, 0.650 in the Netherlands) is omitted from the panels for readability.} Right panels: boxplot of the estimated scale parameter values for all individuals in our data. Source: Statbel, CBS, and authors' calculations.}
    \label{fig:scaleBE}
\end{figure}

While the overall tail behavior appears similar across individuals, substantial heterogeneity arises in the scale parameter. Figure~\ref{fig:scaleBE} shows the estimated covariate effects.

For a constant shape parameter, the scale parameter is directly related to the maximum attainable lifespan (see \eqref{eq:maxlifespan}): higher values correspond to larger maximum lifespans, and lower values to smaller maximum lifespans. It is important to note that these effects are conditional on surviving to the threshold age of 100, and that all covariates are recorded at this age. \added{Since the covariates capture the individual's situation at age 100, they partly reflect health status at that age as well as socio-demographic circumstances. As a consequence, we should interpret the estimated effects as associations rather than causations.}

Several patterns emerge consistently across both countries. We draw the following conclusions for the socio-demographic covariates:
\begin{itemize}
    \item[\texttt{civ}:] Considering civil status, individuals at age 100 who are married, have a partner, or have never been married or been in a partnership (i.e., single) show positive effects on the scale parameter, implying a larger maximum lifespan compared to widowed individuals (the reference category). Divorced individuals show a scale parameter similar to that of widowed individuals, with no significant difference.
    \item[\texttt{hht}:] Household type exhibits more pronounced differences. Using collective (institutional) household as the reference category, we estimate a high and significant value for the scale parameter for individuals living alone in both Belgium and the Netherlands, indicating that these individuals tend to reach among the longest maximum lifespans. In addition, in the Netherlands, an even larger scale parameter is estimated for individuals living in the `other' household type. This category includes private households that do not fit standard family structures, such as individuals living with other relatives, non-family members, or in shared living arrangements. Finally, we also find that people living in family households with children, as well as couples without children, have a significantly higher scale parameter, and thus a higher maximal lifespan, compared to individuals living in collective or institutional households, although the estimated scale remains clearly lower than for individuals living alone. \added{These differences should not necessarily be interpreted as causal. Household type at very advanced ages is likely to be correlated with underlying health status. For example, individuals who live alone are generally more independent, whereas residence in a collective or institutional household often reflects greater care needs or frailty. The estimated effects may therefore reflect selection into household types as well as an effect of the living arrangement itself.}
    \item[\texttt{org}:] Regarding origin, the results differ between Belgium and the Netherlands. In Belgium, individuals with Western European or other non-native backgrounds show higher maximum lifespans compared to the native reference category. The effect is particularly large for individuals of non-Western European origin, which may reflect the very small sample size for this group (146 individuals). In the Netherlands, the estimated effects are much more modest: no significantly different scale parameter is found for individuals with a Western European origin, while a more moderate, but still significant, positive scale parameter is estimated for individuals of non-Western European origin. Given the larger proportion of individuals with a non-Western European origin in the Dutch microdata, we place greater confidence in these estimates.
    \item[\texttt{sex}:] Sex shows a clear and significant effect in both Belgium and the Netherlands: males are associated with a lower scale parameter and, consequently, a shorter maximum lifespan than females.
\end{itemize}

Overall, we conclude that socio-demographic characteristics remain strongly associated with variation in extreme lifespans, even conditional on survival to age 100. \added{These estimated effects are robust: including education in the model or restricting the data to the pre-pandemic period (1995--2019) leaves the estimated scale parameters largely unchanged (Appendices~\ref{app:sensedu} and~\ref{app:senscovid}).}

\subsection{Heterogeneity in the upper endpoint}

\begin{figure}[!ht]
    \centering
    \includegraphics[width=0.45\linewidth]{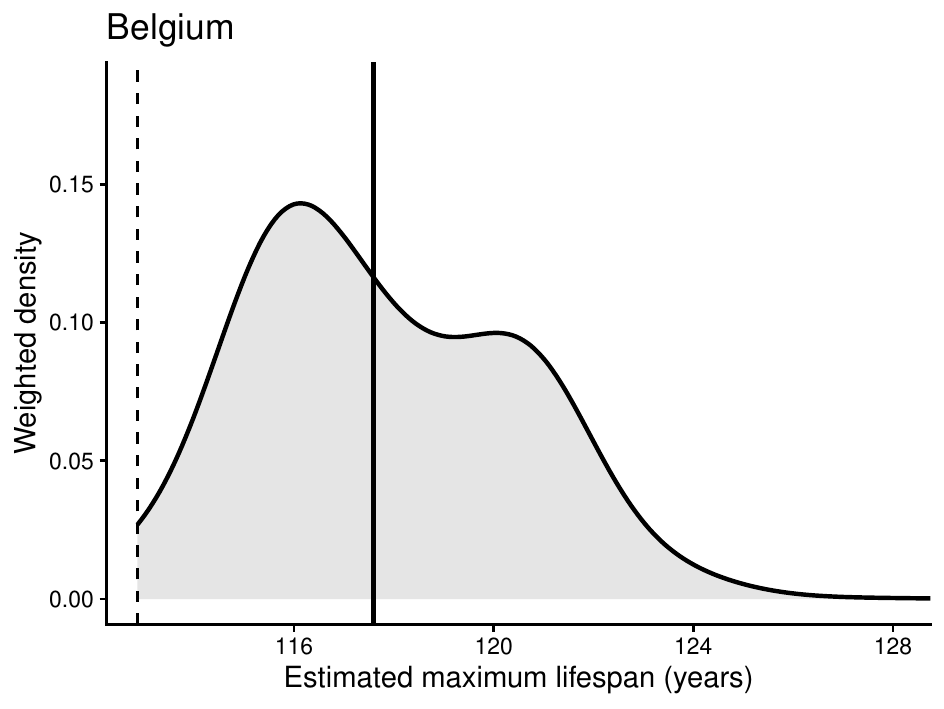}
    \includegraphics[width=0.45\linewidth]{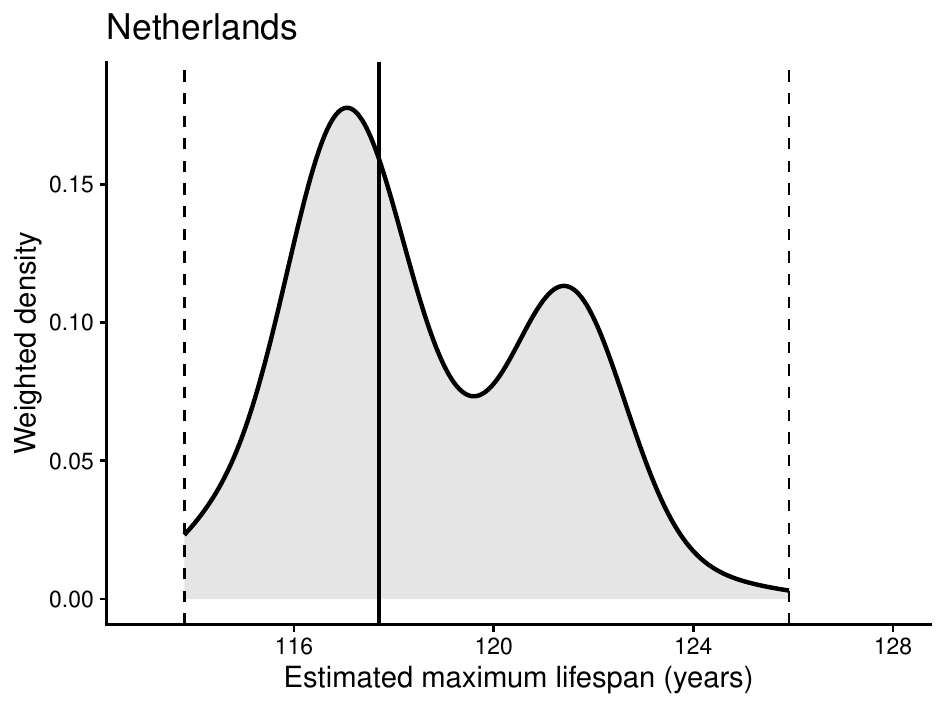}
    \caption{\added{Density of the estimated maximum lifespans across socio-demographic profiles with a frequency greater than 10 in the data set for Belgium (left) and the Netherlands (right). The density is weighted by the observed frequency of each profile in the respective dataset. Dashed vertical lines indicate the minimum and maximum estimated lifespans, while the solid vertical line denotes the median. Source: Statbel, CBS, and authors' calculations.}}
    \label{fig:heteroBENL}
\end{figure}

To quantify the implications of these differences, we estimate the implied maximum lifespan for each observed socio-demographic profile. Figure~\ref{fig:heteroBENL} shows the distribution of these estimates.

We find substantial heterogeneity across socio-demographic profiles. Median estimated maximum lifespans are very similar across countries (Belgium: \added{117.6} years; Netherlands: \added{117.7} years), but the spread is considerable. In Belgium, the range is wider compared to the Netherlands, but this is partly driven by a small number of profiles with a non-Western European origin that have high estimated values. Excluding such extremes leads to very comparable distributions in both countries.

The density is also clearly bimodal. This largely reflects differences between a large group of institutionalized and non-institutionalized individuals, and again highlights that the upper endpoint of the lifespan distribution is not well captured by a single population-wide value.

Focusing on empirically relevant profiles, Table~\ref{tab:mlspanBENL} reports estimated maximum lifespans for the most frequent profiles among individuals aged 100+. Even within this restricted set, differences \added{of up to 10 years} are observed between profiles. This indicates that meaningful heterogeneity persists even among well-represented subpopulations.

\begin{table}[!ht]
\centering
\adjustbox{max width=\textwidth}{%
\begin{tabular}[t]{llllcccc}
\toprule
civ & hht & org & sex & BE freq & NL freq  & BE lifespan & NL lifespan \\
\midrule
widowed   & collective    & native & female & 7\ 416 & 9\ 237 & 115.64 & 116.83 \\
widowed   & single-person & native & female & 5\ 344 & 6\ 821 & 120.68 & 121.45 \\
widowed   & family        & native & female & 1\ 207 & 790    & 118.81 & 119.64 \\
single    & collective    & native & female & 1\ 199 & 1\ 366 & 117.60 & 117.70 \\
widowed   & single-person & native & male   & 1\ 175 & 1\ 534 & 117.01 & 118.35 \\
widowed   & couple        & native & female & 960    & 556    & 117.95 & 118.07 \\
widowed   & collective    & native & male   & 844    & 1\ 476 & 112.86 & 114.40 \\
widowed   & other         & native & female & 798    & 13     & 117.91 & 124.65 \\
single    & single-person & native & female & 465    & 720    & 123.27 & 122.56 \\
married   & couple        & native & male   & 348    & 330    & 116.32 & 117.31 \\
\bottomrule
\end{tabular}}
\caption{\added{Estimated maximum lifespan in Belgium and the Netherlands for the ten most frequently occurring socio-demographic profiles in the Belgian microdata, restricted to individuals whose observed (possibly censored) lifetime exceeds 100 years. Confidence intervals are reported in Appendix~\ref{app:CIs} (Table \ref{tab:mlspanBENLCI}). Source: Statbel, CBS, and authors' calculations. \label{tab:mlspanBENL}}}
\end{table}

\subsection{Magnitude of socio-demographic differences}

To better interpret these differences, Table~\ref{tab:covcontr} evaluates how changes in individual characteristics affect the estimated maximum lifespan relative to a baseline profile. We select as baseline \added{the socio-demographic profile that leads to the lowest estimated maximum lifespan, i.e., a male, native individual who is widowed and resides in a collective household at age 100.}

\begin{table}
\centering
\begin{tabular}[t]{llcc}
\toprule
Type & Level & Lifespan increase (BE) & Lifespan increase (NL) \\
\midrule
civ & divorced      & 0.920  & -0.603 \\
civ & married       & 1.359  & 1.716 \\
civ & single        & 1.606  & 0.743 \\
hht & other         & 1.864  & 6.685 \\
hht & couple        & 1.900  & 1.062 \\
org & west-europe   & 2.257  & 0.293 \\
hht & family        & 2.604  & 2.402 \\
sex & female        & 2.780  & 2.429 \\
hht & single-person & 4.145  & 3.949 \\
org & non-western   & 18.771 & 2.048 \\
\bottomrule
\end{tabular}
\caption{\added{Increase in estimated maximum lifespan when changing one of the socio-demographic characteristics of the reference person: a male, native individual that, at the age of 100, lives in a collective household and is widowed. Confidence intervals are reported in Appendix~\ref{app:CIs} (Table~\ref{tabA:covcontr}). Source: Statbel, CBS, and authors' calculations. \label{tab:covcontr}}}
\end{table}

The largest effects are associated with household composition. Relative to the baseline, living alone \added{is associated with an estimated maximum lifespan that is approximately four years higher in both countries}, while family and couple households are associated with increases of around two years. Sex differences are also substantial, with females exhibiting increases of approximately 2.5 to 3 years. Civil status contributes more modestly, with being married or \added{single} associated with gains of about one to two years relative to widowhood. Lastly, estimates for origin are sensitive to sparse observations, particularly in Belgium. 

Taken together, these results show that the upper endpoint of the lifespan distribution varies systematically across socio-demographic groups. The magnitude of these differences is non-negligible, suggesting that inequalities persist even at the most advanced ages.

\section{Conclusion} \label{sec:conclusion}

In this paper, we revisited the long-standing debate on the existence of an upper limit to the human lifespan by explicitly accounting for socio-demographic heterogeneity at the oldest-old ages. Using population-wide individual-level microdata for Belgium and the Netherlands covering all residents aged 90 and older between 1995 and 2022, we applied an extreme value theory framework that accounts for left truncation and right censoring. Across both countries, \added{the estimated shape parameter is negative, which implies a finite upper endpoint of the fitted lifespan distribution. We note that this is a statement about the tail behaviour of the lifetime distribution based on extreme value theory. These results are consistent with the idea of a biologically fixed limit of human life for the considered populations.} 

Beyond the existence of a finite lifespan limit, our main finding is that the upper tail of the lifespan distribution \added{is not well described by a single population-wide maximum age in Belgium and the Netherlands.} Instead, we document substantial heterogeneity in the estimated upper endpoint across socio-demographic groups, even when conditioning on survival to very old ages. Differences in civil status, household type, sex, and origin are all associated with meaningful variation in the estimated maximum lifespan. These differences persist among profiles that are well represented in the data, indicating that they are not driven solely by sparse observations at the extremes. 

\added{We further note that the observed associations are not necessarily causal. For instance, an important predictor for the estimated upper limit to human life is household type at age 100, a socio-demographic covariate which partly reflects underlying health and frailty. As a result, the estimated heterogeneity likely reflects both socio-demographic differences and health-related selection at the oldest ages.}

In the present analysis, heterogeneity is captured through the scale parameter of the generalized Pareto distribution, while imposing a common extreme value index across groups. This specification ensures a shared tail behavior across the population, \added{and is supported by our data, since allowing the extreme value index to vary across socio-demographic groups did not improve the model fit (Appendix~\ref{app:sensshape}). Still, more flexible specifications could be explored in future work.} In addition, although the data are population-wide, some socio-demographic groups, particularly those defined by migration background, remain sparsely represented at the highest ages, and corresponding estimates should therefore be interpreted with caution. \added{Our analysis also does not separate calendar-period and cohort effects. Because observation ends in 2022, only the earliest cohorts are observed at the most extreme ages, so cohort-specific tail behaviour cannot be identified from these data. For calendar time, \citet{einmahl2019limits} find no significant trend in the estimated endpoint for the Netherlands.} \added{Finally, our results are based on data from only Belgium and the Netherlands. Whether comparable socio-demographic heterogeneity arises in other populations remains a question for future research.}

Overall, our results show that inequalities in longevity persist even at the most advanced ages and are reflected in the upper tail of the lifespan distribution. Moving beyond a single-number characterization of human lifespan provides a more nuanced perspective on extreme longevity and contributes to a better understanding of the social stratification underlying the statistical limits of human life.

\section*{Acknowledgements}
This study is part of the research programme at the Research Centre for Longevity Risk—a joint initiative of NN Group and the University of Amsterdam, with additional funding from the Dutch government's Public Private Partnership programme.

\section*{Data availability}
\added{Results for the Netherlands are based on calculations by the Research Centre for Longevity Risk (RCLR) in project number 3061 using non-public microdata from Statistics Netherlands, covering 1995--2022. The analysis draws on the population and death registers (\texttt{Gbapersoontab}, \texttt{Gbaoverlijdentab}), civil status (\texttt{Gbaburgerlijkestaatbus}), household composition (\texttt{Gbahuishoudensbus}), the parent--child links (\texttt{Gbakindbus}, \texttt{Kindoudertab}), and educational attainment (\texttt{Hoogsteopltab}), including that of the highest-educated child, used when own education is missing. Under certain conditions these microdata are accessible for statistical and scientific research: \href{https://www.cbs.nl/en-gb/our-services/customised-services-microdata/microdata-conducting-your-own-research}{Microdata: Conducting your own research | CBS}.}

\added{Results for Belgium are based on calculations by the Research Centre for Longevity Risk (RCLR) using non-public microdata from Statbel (the Belgian statistical office), covering 1995--2022. These microdata are constructed from the population register (\texttt{TF\_Stock}), the death register (\texttt{TF\_Death}), and the 2001, 2011, and 2021 censuses. From these sources, we use sex (\texttt{CD\_SEX}), date of birth (\texttt{DT\_BTH}), civil status (\texttt{CD\_CIV}), origin (\texttt{CD\_DSCNT}), household information (\texttt{ID\_HH}, \texttt{HH\_TYPE\_LIPRO}, \texttt{HH\_POS\_LIPRRO}), exact age at death, together with educational attainment from the censuses. Under certain conditions these microdata are accessible for statistical and scientific research: \href{https://statbel.fgov.be/en/about-statbel/what-we-do/microdata-research}{Microdata for research | Statbel}.}

\added{The \texttt{R}-code used to produce all results for Belgium and the Netherlands, including the model fitting, covariate selection, and the sensitivity and diagnostic analyses, is publicly available in the GitHub repository: \url{https://github.com/jensrobben/lifespan-evt}. This code operates on the CBS and Statbel microdata, which are not included. Researchers with approved access to these microdata can reproduce the results.}

{
\bibliography{plos_bibtex_sample}
}

\appendix
\setcounter{figure}{0}
\setcounter{table}{0}
\renewcommand{\thefigure}{A\arabic{figure}}
\renewcommand{\thetable}{A\arabic{table}}
\newpage
{\LARGE \bfseries Appendix}
\section{Validation of dates of birth and death} \label{app:validation}
\added{In the Belgian microdata (Statbel), dates of birth and death are taken from the administrative register of natural persons. The date of death is the date stated on the death certificate by the physician who officially certified the death. Statbel does not carry out a validation of these dates themselves beyond verifying that they are valid calendar dates. Invalid dates are adjusted to the nearest date within the same month. In principle the year of birth is known, but the exact date is not always fully observed. In those cases, the day of birth is set to the 16th of the month if only the day is missing, and to 1st of July if both the day and the month are missing.}

\added{In the Dutch microdata (CBS), dates of birth and death are registered by the municipalities in the Personal Records Database (BRP). The BRP Act specifies which documents individuals must submit upon registration (Articles 2.8, 2.15, and 2.17). In the absence of any documentation individuals can make a declaration under oath. Individuals who are not yet known to the register must appear in person to verify their identity. For births in the Netherlands a statement from a doctor or midwife may be requested. In a limited number of cases the date of birth is only partially known, for instance because it does not appear on any document or because the individual does not know the exact date. This mainly concerns individuals born in countries with less complete civil registration, for whom the declared date is the best available information. Partially missing dates are imputed: the 1st of July if only the year of birth is known, and the 15th of the month if the day of birth is unknown.}

\section{Maximum lifespan estimation using the generalized Pareto model} \label{app:theory}
\subsection{Extreme value theory framework} \label{app:evt}
To study the upper tail of the lifespan distribution, we rely on extreme value theory (EVT). Let $X$ be a random variable that denotes the age at death of a particular individual. Under some mild regularity conditions as stated in the Pickands-Balkema-de Haan theorem \citep{pickands1975statistical, balkema1974residual}, the conditional distribution of threshold lifetime exceedances,
\begin{align*}
Y = X - u,
\end{align*}
given $X > u$, can be approximated by the Generalized Pareto Distribution (GPD) for a sufficiently high threshold age $u$:  
\begin{equation} \label{eq:gpapprox}
\begin{aligned}
\Pr(Y \le y \mid X > u) \approx
\begin{cases}
1 - \left(1 + \xi \dfrac{y}{\sigma}\right)_+^{-1/\xi}, & \xi \neq 0,\\[1em]
1 - \exp\left(-\dfrac{y}{\sigma}\right), & \xi = 0, \; y \ge 0,
\end{cases}   
\end{aligned}
\end{equation}
where $\sigma > 0$ is a scale parameter and $\xi$ is the shape parameter or extreme value index (EVI). The EVI $\xi$ determines the tail behavior of the distribution: $\xi > 0$ corresponds to a heavy-tailed (Pareto-type) distribution with unbounded lifespan, $\xi = 0$ corresponds to an exponentially decaying tail (Gumbel class), and $\xi < 0$ corresponds to a light-tailed distribution with a finite upper endpoint $y_{\max}$, which reflects the maximum attainable age implied by the EVT framework:
\begin{align} \label{eq:maxL}
x^* = u + y_{\max} = u - \frac{\sigma}{\xi}.   
\end{align}
In other words, the EVI $\xi$ determines whether this model for the future lifetime of individuals implies a finite maximum lifetime ($\xi < 0$) that no individual will exceed, or does not impose such a limit ($\xi \geq 0$).

\subsection{Accounting for censoring and truncation in the likelihood} \label{app:likelihood}
As mentioned earlier, the individual-level lifespan microdata are subject to both right censoring and left truncation. Right censoring arises for individuals who emigrate or who are still alive at the end of the observation period (year 2022), in which case only a lower bound on the age at death is observed. Left truncation occurs because individuals enter the study at different ages, and only those who have survived beyond their entry age are included in the sample. In particular, for older birth cohorts, individuals may enter observation at ages well above the selected threshold age $u$, while individuals from the same cohorts who died between $u$ and the entry age are unobserved. As a result, inference on extreme lifespans must be conducted conditionally on delayed entry and censoring, requiring explicit adjustment for in the likelihood.

Let $t_i$ denote the age at which individual $i$ enters observation. For any chosen threshold age $u$, the age $t_i$ will be either $t_i=u$ if the life is observed on its $u$-th birthday, or $t_i>u$ if the life is already older than the threshold age when it is first observed (left truncation).
We denote by $x_i$ the age at death if observed, and $c_i$ the age at right censoring if alive at the end of the observation period (year 2022) or whenever the life is censored. 

We define $y_i = \max(x_i, c_i) - u$ and $\delta_i$ the event indicator: $\delta_i = 1$ if death is observed ($x_i \leq c_i$) and $\delta_i = 0$ if the observation is right censored ($x_i > c_i$). The likelihood contribution for individual $i$ with threshold exceedance is:  
\begin{equation} \label{eq:loglikcontr}
L(\beta_0,\boldsymbol{\beta}_\sigma, \xi \mid \mathbf{z}_i, t_i, y_i) =
\begin{cases}
\displaystyle\frac{f_{\text{gpd}}(y_i; \sigma_i, \xi)}{S_{\text{gpd}}(t_i - u; \sigma_i, \xi)}, & \delta_i = 1, \\[12pt]
\displaystyle\frac{S_{\text{gpd}}(y_i; \sigma_i, \xi)}{S_{\text{gpd}}(t_i - u; \sigma_i, \xi)}, & \delta_i = 0,
\end{cases}
\end{equation}
where $f_{\text{gpd}}$ and $S_{\text{gpd}}$ denote the density and survival function of the generalized Pareto distribution, and the denominator accounts for left truncation. Note that $t_i-u = 0$ for all lifetimes that are not left truncated. In that case, the denominator in \eqref{eq:loglikcontr} is $S_{\text{gpd}}(t_i - u; \sigma_i, \xi) = 1$.

Under the assumption that, conditional on covariates $\mathbf{z}_i$, the exceedances $Y_i$ are independent across individuals and that the age at death is independent of the truncation and censoring mechanisms (i.e., independent left truncation and non-informative right censoring), the full log-likelihood is:
\begin{align} \label{eq:loglik}
\ell(\beta_0,\boldsymbol{\beta}_\sigma, \xi)
= \sum_{\substack{i: y_i > 0}} \log L(\beta_0, \boldsymbol{\beta}_\sigma, \xi \mid \mathbf{z}_i, t_i, y_i).  
\end{align}
The parameters $(\beta_0,\boldsymbol{\beta}_\sigma, \xi)$ are estimated using maximum likelihood estimation (MLE), and standard errors are computed using the observed Fisher information matrix. In the sequel, we assume that the intercept $\beta_0$ is included in the coefficient vector $\boldsymbol{\beta}_\sigma$ and that the covariate vector $\boldsymbol{z}_i$ is augmented with a leading constant 1.

\subsection{Full log-likelihood with covariates} \label{app:fullloglik}
The full log-likelihood for the Generalized Pareto model with covariate-dependent scale $\sigma_i = \exp(\boldsymbol{\beta}_\sigma^T \mathbf{z}_i)$ and constant extreme value index $\xi \neq 0$, accounting for left truncation at age $t_i$ and right censoring at age $c_i$, can be written as:
\begin{equation} \label{eqref:fullloglik}
\ell(\boldsymbol{\theta}) = \sum_{i: y_i>0} 
\Bigg[ - \delta_i \, \boldsymbol{\beta}_\sigma^T \mathbf{z}_i 
- \left(\frac{1}{\xi} + \delta_i \right) \log \Big(1 + \xi \, y_i \, e^{-\boldsymbol{\beta}_\sigma^T \mathbf{z}_i} \Big) 
+ \frac{1}{\xi} \log \Big(1 + \xi \, a_i \, e^{-\boldsymbol{\beta}_\sigma^T \mathbf{z}_i} \Big) 
\Bigg],
\end{equation}
where $\boldsymbol{\theta} = (\boldsymbol{\beta}_{\sigma}, \xi)$, $y_i = \max(x_i, c_i) - u$ denotes the exceedance over the threshold age $u$, $a_i = t_i - u$ denotes the exceedance of the entry age above the threshold age $u$, and $\delta_i$ is the event indicator. 

\subsection{Maximum likelihood estimation and gradients} \label{app:mle}
We estimate the parameter vector $\boldsymbol{\theta}$ via maximum likelihood by numerically maximizing the full log-likelihood in~\eqref{eqref:fullloglik}. We use the BFGS optimization algorithm, which is a quasi-Newton method that benefits from the availability of the first-order derivatives of the log-likelihood with respect to $\boldsymbol{\theta}$. The gradient of the log-likelihood with respect to the regression coefficients $\boldsymbol{\beta}_\sigma$ equals:
\begin{equation}
\frac{\partial \ell(\boldsymbol{\theta})}{\partial \boldsymbol{\beta}_\sigma} = \sum_{i: y_i>0} 
\Bigg[ -\delta_i + \left(\frac{1}{\xi} + \delta_i\right) \frac{\xi y_i e^{-\boldsymbol{\beta}_\sigma^T \mathbf{z}_i}}{1 + \xi y_i e^{-\boldsymbol{\beta}_\sigma^T \mathbf{z}_i}} 
- \frac{\xi a_i e^{-\boldsymbol{\beta}_\sigma^T \mathbf{z}_i}}{1 + \xi a_i e^{-\boldsymbol{\beta}_\sigma^T \mathbf{z}_i}} \Bigg] \mathbf{z}_i,
\end{equation}
and the derivative with respect to the extreme value index $\xi$ equals:
\begin{equation}
\frac{\partial \ell(\boldsymbol{\theta})}{\partial \xi} = \sum_{i: y_i>0} 
\Bigg[ - \left(\frac{1}{\xi} + \delta_i\right) \frac{y_i e^{-\boldsymbol{\beta}_\sigma^T \mathbf{z}_i}}{1 + \xi y_i e^{-\boldsymbol{\beta}_\sigma^T \mathbf{z}_i}} 
+ \frac{1}{\xi^2} \log \left( \frac{1 + \xi y_i e^{-\boldsymbol{\beta}_\sigma^T \mathbf{z}_i}}{1 + \xi a_i e^{-\boldsymbol{\beta}_\sigma^T \mathbf{z}_i}} \right)
+ \frac{1}{\xi} \frac{a_i e^{-\boldsymbol{\beta}_\sigma^T \mathbf{z}_i}}{1 + \xi a_i e^{-\boldsymbol{\beta}_\sigma^T \mathbf{z}_i}} \Bigg].
\end{equation}
These analytical gradients are used in the BFGS algorithm to ensure better convergence of the maximum likelihood estimation.

\subsection{Confidence intervals via the observed Fisher information} \label{app:fisherCI}
Next, we construct confidence intervals for the parameter vector $\boldsymbol{\theta} = (\boldsymbol{\beta}_\sigma, \xi)$ using the observed Fisher information matrix. Under standard maximum likelihood theory, we can approximate the asymptotic covariance matrix of $\hat{\boldsymbol{\theta}}$ by the inverse of the negative Hessian of the log-likelihood evaluated at the maximum likelihood estimates:
\begin{equation}
\widehat{\text{Cov}}(\hat{\boldsymbol{\theta}}) \approx \mathcal{I}(\hat{\boldsymbol{\theta}})^{-1}, \quad \mathcal{I}(\hat{\boldsymbol{\theta}}) = - \frac{\partial^2 \ell(\boldsymbol{\theta})}{\partial \boldsymbol{\theta} \partial \boldsymbol{\theta}^T} \Bigg|_{\boldsymbol{\theta} = \hat{\boldsymbol{\theta}}}.
\end{equation}
Approximate $100(1-\alpha)\%$ confidence intervals for each parameter $\theta_j$ are then obtained as:
\begin{equation}
\hat{\theta}_j \pm z_{\alpha/2} \, \text{SE}(\hat{\theta}_j), \quad \text{where } \text{SE}(\hat{\theta}_j) = \sqrt{\widehat{\text{Var}}(\hat{\theta}_j)},
\end{equation}
where $z_{\alpha/2}$ is the $(1-\alpha/2)$-quantile of the standard normal distribution. We numerically approximate the Hessian matrix of the log-likelihood through the BFGS routine.

\subsection{Delta method for the maximum lifespan} \label{app:delta}
Conditionally on the extreme value index being negative, $\xi < 0$, the maximum lifespan for an individual with covariate vector $\boldsymbol{z}_i$ is given by:
\begin{align*}
    g(\boldsymbol{\theta}) = u - \frac{\exp(\boldsymbol{\beta}_\sigma^T \boldsymbol{z}_i)}{\xi}.
\end{align*}
To quantify the uncertainty of the estimated maximum lifespan $\hat{g}(\hat{\boldsymbol{\theta}})$, we apply the delta method. Specifically, if $\hat{\boldsymbol{\theta}}$ is asymptotically normal with covariance matrix $\widehat{\text{Cov}}(\hat{\boldsymbol{\theta}})$, then:
\begin{align}
    \text{Var}(\hat{g}(\hat{\boldsymbol{\theta}})) \approx \nabla_{\boldsymbol{\theta}} g(\hat{\boldsymbol{\theta}})^T \, \widehat{\text{Cov}}(\hat{\boldsymbol{\theta}}) \, \nabla_{\boldsymbol{\theta}} g(\hat{\boldsymbol{\theta}}),
\end{align}
where $\nabla_{\boldsymbol{\theta}} g(\boldsymbol{\theta})$ denotes the gradient of $g(\boldsymbol{\theta})$ with respect to $\boldsymbol{\theta} = (\boldsymbol{\beta}_\sigma, \xi)$.

The derivatives of $g(\boldsymbol{\theta})$ are:
\begin{align}
    \frac{\partial g(\boldsymbol{\theta})}{\partial \boldsymbol{\beta}_\sigma} &= - \frac{\exp(\boldsymbol{\beta}_\sigma^T \boldsymbol{z}_i)}{\xi} \, \boldsymbol{z}_i, \\
    \frac{\partial g(\boldsymbol{\theta})}{\partial \xi} &= \frac{\exp(\boldsymbol{\beta}_\sigma^T \boldsymbol{z}_i)}{\xi^2}.
\end{align}

Consequently, the approximate $100(1-\alpha)\%$ confidence interval for the maximum lifespan is:
\begin{align}
    g(\hat{\boldsymbol{\theta}}) \pm z_{\alpha/2} \, \sqrt{\nabla_{\boldsymbol{\theta}} g(\hat{\boldsymbol{\theta}})^T \, \widehat{\text{Cov}}(\hat{\boldsymbol{\theta}}) \, \nabla_{\boldsymbol{\theta}} g(\hat{\boldsymbol{\theta}})}.
\end{align}

\subsection{Limitations of the delta method and bootstrap alternative} \label{app:bootstrapCI}
While the delta method provides a convenient analytical approximation for the variance of the estimated maximum lifespan, it is not always ideal in the context of extreme value theory. This is because the delta method relies on a linear approximation of $g(\boldsymbol{\theta})$ around the maximum likelihood estimates and assumes asymptotic normality of $\hat{\boldsymbol{\theta}}$ \citep{davison1990models}. In practice, for small sample sizes of exceedances or when the estimated extreme value index $\xi$ is close to zero, these assumptions may be violated, potentially leading to inaccurate confidence intervals. However, in our study, we have more than 20\,000 exceedances over the threshold age $u$, which provides a sufficiently large sample for the asymptotic approximations to be more reliable.

An alternative approach to quantify uncertainty is the non-parametric bootstrap. In this framework, we repeatedly sample with replacement from the observed exceedances, refit the generalized Pareto model to each resampled dataset, and recompute the derived maximum lifespan $g(\boldsymbol{\theta})$. This generates an empirical distribution of $g$ from which confidence intervals can be obtained. While computationally more intensive than the delta method, the bootstrap does not rely on asymptotic normality and is particularly useful when the sample size is moderate or the model exhibits strong nonlinearity.

\subsection{Results} \label{app:CIs}
We estimate the model on the Belgian and Dutch microdata, which include individuals who died or are still alive beyond the threshold age of 100 years during 1995–2022. Table~\ref{tabA:cis}  presents the estimated socio-demographic effects on the scale parameter along with the corresponding $95\%$ confidence intervals, computed using both the observed Fisher information matrix (Appendix~\ref{app:fisherCI}) and the non-parametric bootstrap approach (Appendix~\ref{app:bootstrapCI}). The confidence intervals obtained from both methods are very similar. For the shape parameter, we obtained a point estimate of \added{-0.1354 in Belgium (CI: [-0.1455, -0.1253], Bootstrap CI: [-0.1560, -0.1221])} and \added{-0.1138 in the Netherlands (CI: [-0.1245, -0.1030], Bootstrap CI: [-0.1259,-0.1032])}. We note that the confidence intervals for the shape parameter are slightly wider when using the bootstrap approach.

\begin{table}[htbp]
\centering
\adjustbox{max width = \textwidth}{
\begin{tabular}{lccc @{\hspace{0.75cm}} ccc}
\toprule
& \multicolumn{3}{c}{\textbf{Belgium}} &  \multicolumn{3}{c}{\textbf{The Netherlands}}\\
 & Estimate & 95\% CI & 95\% Boot.~CI & Estimate & 95\% CI & 95\% Boot.~CI\\
 \midrule
\textbf{Intercept} & 0.750$^{*}$ & (0.727, 0.774) & (0.728, 0.777) & 0.650$^{*}$ & (0.629, 0.671) & (0.629, 0.672)\\
\addlinespace
\multicolumn{7}{l}{\textbf{Civil status}}\\
Single & 0.118$^{*}$ & (0.076, 0.160) & (0.076, 0.160) & 0.050$^{*}$ & (0.012, 0.088) & (0.013, 0.088)\\
Married & 0.100$^{*}$ & (0.017,0.184) & (0.016,0.180) & 0.113$^{*}$ & (0.032, 0.193) & (0.031, 0.191)\\
Divorced & 0.069 & (-0.019,0.157) & (-0.017,0.152) & -0.043 & (-0.110, 0.024) & (-0.104, 0.017)\\
\addlinespace
\multicolumn{7}{l}{\textbf{Household type}}\\
Single-person & 0.279$^{*}$ & (0.251,0.308) & (0.251,0.309) & 0.242$^{*}$ & (0.218, 0.266) & (0.218, 0.266)\\
Couple & 0.138$^{*}$ & (0.088,0.188) & (0.083,0.196) & 0.071$^{*}$ & (0.012, 0.130) & (0.011, 0.131) \\
Family & 0.184$^{*}$ & (0.135,0.233) & (0.133,0.234) & 0.154$^{*}$ & (0.100, 0.208) & (0.099, 0.209) \\
Other & 0.135$^{*}$ & (0.077,0.193) & (0.077,0.191) & 0.381$^{*}$ & (0.126, 0.637) & (0.062, 0.672) \\
\addlinespace
\multicolumn{7}{l}{\textbf{Origin}}\\
West-Europe & 0.162$^{*}$ & (0.086,0.238) & (0.074,0.246) & 0.020 & (-0.028, 0.068) & (-0.026, 0.065)\\
Non-western & 0.900$^{*}$ & (0.669,1.131) & (0.675,1.141) & 0.133$^{*}$ & (0.074, 0.192) & (0.072, 0.192)\\
\addlinespace
\multicolumn{7}{l}{\textbf{Sex}}\\
Male & -0.196$^{*}$ & (-0.232,-0.160) & (-0.236,-0.157) & -0.156$^{*}$ & (-0.187, -0.125) & (-0.186, -0.125)\\
\midrule
\bottomrule
\end{tabular}}
\caption{\added{Estimated effects in the scale parameter with 95$\%$ confidence intervals based on the observed Fisher information matrix (95$\%$ CI) and the non-parametric bootstrap (95$\%$ bootstrap CI). Asterisks indicate estimates for which both confidence intervals exclude zero. Source: Statbel, CBS, and authors' calculations. \label{tabA:cis}}}
\end{table}

Next, we compute confidence intervals for the maximum lifespan using both approaches for the 10 most frequently occurring profiles in the microdata. The estimates for these maximum lifespans are presented in Table~\ref{tab:mlspanBENL} of the main text, together with the frequency in the Belgian and Dutch datasets. For reference, Table~\ref{tab:top10} lists the 10 most frequently occurring profiles.

\begin{table}[!ht]
\centering
\adjustbox{max width=\textwidth}{%
\begin{tabular}[t]{cllll}
\toprule
Profile ID & civ  & hht & org & sex  \\
\midrule
ID1  & widowed & collective     & native & female  \\
ID2  & widowed & single-person  & native & female  \\
ID3  & widowed & family         & native & female  \\
ID4  & single  & collective     & native & female  \\
ID5  & widowed & single-person  & native & male  \\
ID6  & widowed & couple         & native & female  \\
ID7  & widowed & collective     & native & male  \\
ID8  & widowed & other          & native & female  \\
ID9  & single  & single-person  & native & female  \\
ID10 & married & couple         & native & male  \\
\bottomrule
\bottomrule
\end{tabular}}
\caption{\added{The ten most frequently occurring socio-demographic profiles in the Belgian and Dutch microdata, restricted to individuals whose observed (possibly censored) lifetime exceeds 100 years.  \label{tab:top10}}}
\end{table}

Table~\ref{tab:mlspanBENLCI} lists the point-estimates for the maximum lifespan for the 10 most occurring profiles as well as the 95$\%$ confidence intervals that are constructed using both the observed Fisher information matrix as well as the non-parametric bootstrap approach. We find that the confidence intervals do not always show overlapping regions between the different profiles, which indicates there is a significant heterogeneity in the estimated maximum lifespans. The confidence intervals computed using the non-parametric bootstrap approach turn out to be slightly wider for Belgium. Since the confidence intervals around the estimated maximum lifespans in Belgium and the Netherlands consistently overlap for the 10 most occurring profiles, we cannot conclude that these lifespans differ significantly between the two countries.

\begin{table}[!ht]
\centering
\adjustbox{max width = \textwidth}{
\begin{tabular}{lccc @{\hspace{0.5cm}} ccc}
\toprule
& \multicolumn{3}{c}{\textbf{Belgium}} &  \multicolumn{3}{c}{\textbf{The Netherlands}}\\
Profile & Lifespan & 95\% CI & 95\% Boot.~CI & Lifespan & 95\% CI & 95\% Boot.~CI\\
 \midrule
ID1  & 115.64 & (114.63,116.66) & (113.80,117.20) & 116.83 & (115.43, 118.24) & (115.40, 118.37) \\
ID2  & 120.68 & (119.24,122.13) & (118.22,122.79) & 121.45 & (119.58, 123.31) & (119.51, 123.53) \\
ID3  & 118.81 & (117.32,120.30) & (116.57,120.76) & 119.64 & (117.75, 121.53) & (117.68, 121.67) \\
ID4  & 117.60 & (116.27,118.92) & (115.49,119.41) & 117.70 & (116.11, 119.30) & (116.10, 119.41) \\
ID5  & 117.01 & (115.75,118.27) & (114.87,118.79) & 118.35 & (116.71, 120.00) & (116.69, 120.13) \\
ID6  & 117.95 & (116.56,119.35) & (115.79,119.90) & 118.07 & (116.25, 119.90) & (116.24, 120.06) \\
ID7  & 112.86 & (111.95,113.78) & (111.25,114.19) & 114.40 & (113.16, 115.65) & (113.16, 115.75) \\
ID8  & 117.91 & (116.38,119.44) & (115.67,119.93) & 124.65 & (118.04, 131.25) & (117.49, 133.17) \\
ID9  & 123.27 & (121.38,125.15) & (120.44,125.76) & 122.56 & (120.44, 124.68) & (120.41, 124.83) \\
ID10 & 116.32 & (114.68,117.97) & (114.12,118.34) & 117.31 & (115.38, 119.24) & (115.48, 119.34) \\
\bottomrule
\bottomrule
\end{tabular}}
\caption{\added{Estimated maximum lifespan for the ten most frequently occurring socio-demographic profiles in the Belgian and Dutch microdata, restricted to individuals whose observed (possibly censored) lifetime exceeds 100 years. We also provide 95$\%$ confidence intervals computed using (1) the observed Fisher information matrix (\texttt{95$\%$ CI}) and (2) the non-parametric bootstrap approach (\texttt{95$\%$ Boot.~CI}). Source: Statbel, CBS, and authors' calculations. \label{tab:mlspanBENLCI}}}
\end{table}

Finally, we compute confidence intervals for the lifespan increases relative to the selected baseline profile (Table~\ref{tabA:covcontr}), again using both approaches to construct 95$\%$ confidence intervals.

\begin{table}[!htb]
\centering
\adjustbox{max width = \textwidth}{
\begin{tabular}{ll ccc @{\hspace{0.5cm}} ccc}
\toprule
& & \multicolumn{3}{c}{\textbf{Belgium}} &  \multicolumn{3}{c}{\textbf{The Netherlands}}\\
Type & Level & Diff. & 95\% CI & 95\% Boot.~CI & Diff. & 95\% CI & 95\% Boot.~CI\\
\midrule
civ & divorced      & 0.920        & (-0.29, 2.13)  & (-0.26, 2.10)  & -0.603 & (-1.53, 0.32) & (-1.42, 0.25) \\
civ & married       & 1.359$^{*}$  & (0.17, 2.54)   & (0.17, 2.56)   & 1.716$^{*}$ & (0.42, 3.01) & (0.44, 3.05)\\
civ & single        & 1.606$^{*}$  & (0.99, 2.22)   & (0.97, 2.23)   & 0.743$^{*}$ & (0.17, 1.32) & (0.18, 1.32) \\
hht & other         & 1.864$^{*}$  & (1.00, 2.72)   & (0.98, 2.71)   & 6.685$^{*}$ & (1.27, 12.10) & (0.90, 13.77) \\
hht & couple        & 1.900$^{*}$  & (1.17, 2.63)   & (1.06, 2.73)   & 1.062$^{*}$ & (0.15, 1.98) & (0.16, 2.01) \\
org & west-europe   & 2.257$^{*}$  & (1.10, 3.41)   & (0.89, 3.61)   & 0.293 & (-0.41, 1.00) & (-0.37, 0.96) \\
hht & family        & 2.604$^{*}$  & (1.84, 3.37)   & (1.77, 3.37)   & 2.402$^{*}$ & (1.49, 3.31) & (1.47, 3.39) \\
sex & female        & 2.780$^{*}$  & (2.27, 3.29)   & (2.14, 3.38)   & 2.429$^{*}$ & (1.92, 2.94) & (1.91, 2.94) \\
hht & single-person & 4.145$^{*}$  & (3.60, 4.69)   & (3.45, 4.78) & 3.949$^{*}$ & (3.39, 4.50) & (3.39, 4.53) \\
org & non-western   & 18.771$^{*}$ & (11.32, 26.22) & (11.73, 27.66) & 2.048$^{*}$ & (1.06, 3.04) & (1.06, 3.10) \\
\bottomrule
\end{tabular}}
\caption{\added{Increase in estimated maximum lifespan when changing one of the socio-demographic characteristics of the reference person: a male, native individual that, at the age of 100, lives in a collective household and is widowed. We also provide 95$\%$ confidence intervals computed using (1) the observed Fisher information matrix (\texttt{95$\%$ CI}) and (2) the non-parametric bootstrap approach (\texttt{95$\%$ Boot.~CI}). Asterisks indicate estimates for which both confidence intervals exclude zero. Source: Statbel, CBS, and authors' calculations. \label{tabA:covcontr}}}
\end{table}

\section{Model diagnostics and sensitivity analyses} \label{app:diagsens}
\subsection{Residual diagnostics for the final covariate model} \label{app:diagnostics}
\added{The Q--Q plots in Section~\ref{subsec:QQ.GPD} assess the generalized Pareto approximation for the pooled fit. To evaluate the goodness-of-fit for the final covariate model, we use Cox--Snell residuals adapted for delayed entry (left truncation) and right censoring  \citep{cox1968general, kalbfleisch2002statistical}. For individual $i$ with estimated scale $\hat{\sigma}_i$, common shape $\hat{\xi}$, entry exceedance $a_i = t_i - u$, and observed exceedance $y_i$, the residual is defined as the negative logarithm of the estimated conditional survival probability:
\begin{equation}
r_i = -\log \left(\dfrac{S_{\text{gpd}}(y_i;\hat{\sigma}_i, \hat{\xi})}{S_{\text{gpd}}(a_i;\hat{\sigma}_i,\hat{\xi})} \right) = \frac{1}{\hat{\xi}} \left[ \log\left(1 + \hat{\xi}\, \frac{y_i}{\hat{\sigma}_i}\right) - \log\left(1 + \hat{\xi}\, \frac{a_i}{\hat{\sigma}_i}\right) \right].
\end{equation}
Delayed entry is accounted for through the conditional survival function, while right-censored observations remain right-censored in the residuals. Under a correctly specified model, the residuals $r_i$ then form an i.i.d.~sample from a standard exponential distribution. We therefore treat $(r_i, \delta_i)$ as a right-censored sample from a standard exponential distribution and estimate its survivor function using the Kaplan--Meier estimator. Under a correct model specification, the estimated cumulative hazard satisfies $\widehat{H}(r) \approx r$, so that the estimate should track the 45-degree line through the origin. We obtain pointwise 95$\%$ confidence bands from the Kaplan--Meier estimator of the survival function of the residuals, using Greenwood's variance formula on the log--log scale. The resulting limits $(S_L(r), S_U(r))$ for the survival function are transformed to the cumulative-hazard scale through the transformation $H(r) = -\log S(r)$, which results in the band ${[-\log S_U(r), -\log S_L(r)]}$.}

\begin{figure}[!ht]
    \centering
    \includegraphics[width=\linewidth]{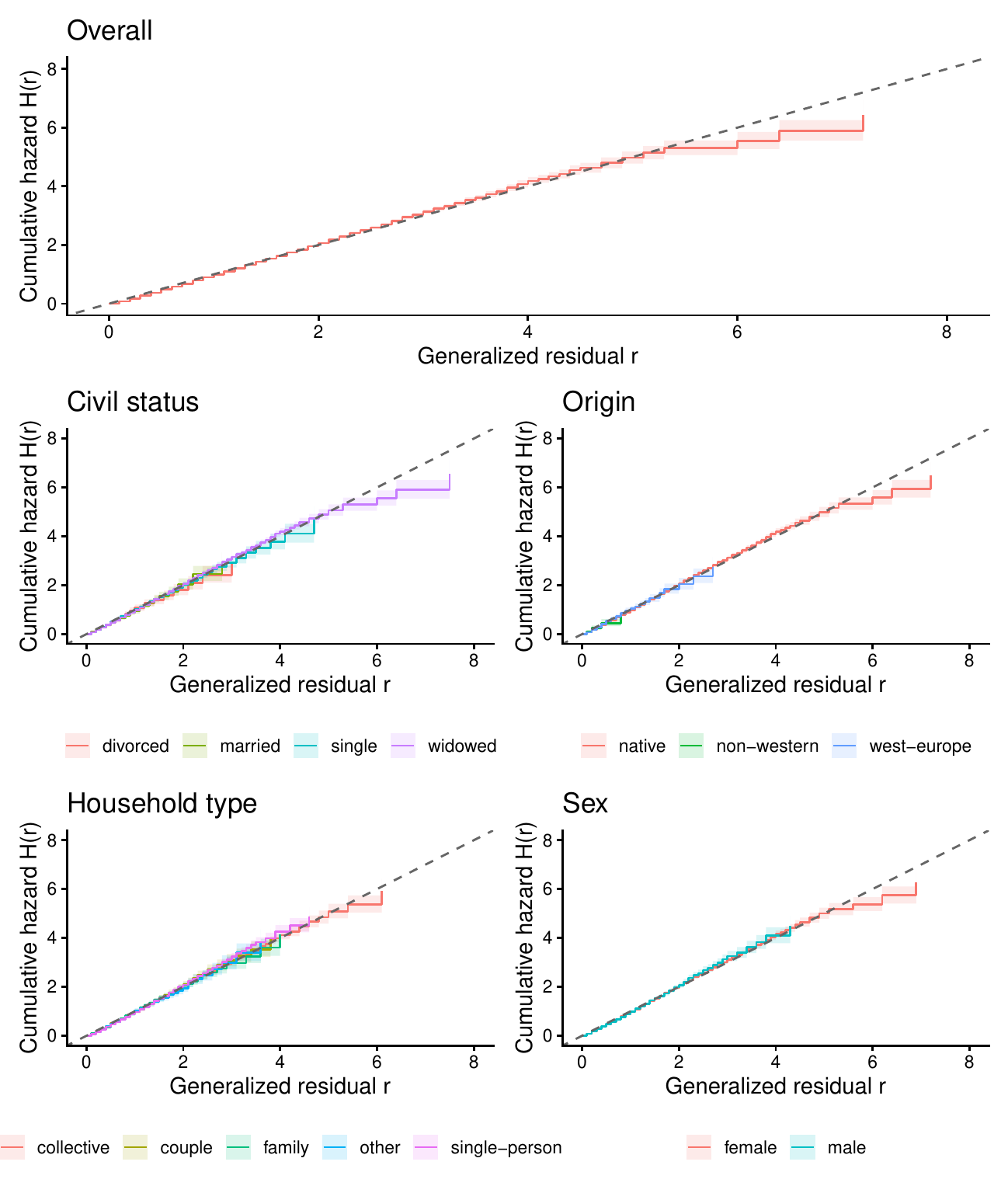}
    \caption{\added{Cumulative hazard diagnostics based on Cox--Snell residuals for the final covariate model in Belgium, overall and stratified by covariates in the final specification. The dashed line indicates the 45-degree reference line implied by a correctly specified model. Shaded areas denote pointwise 95$\%$ confidence bands. Source: Statbel and authors' calculations.}}
    \label{figA:diagnosticsBE}
\end{figure}

\begin{figure}[!ht]
    \centering
    \includegraphics[width=\linewidth]{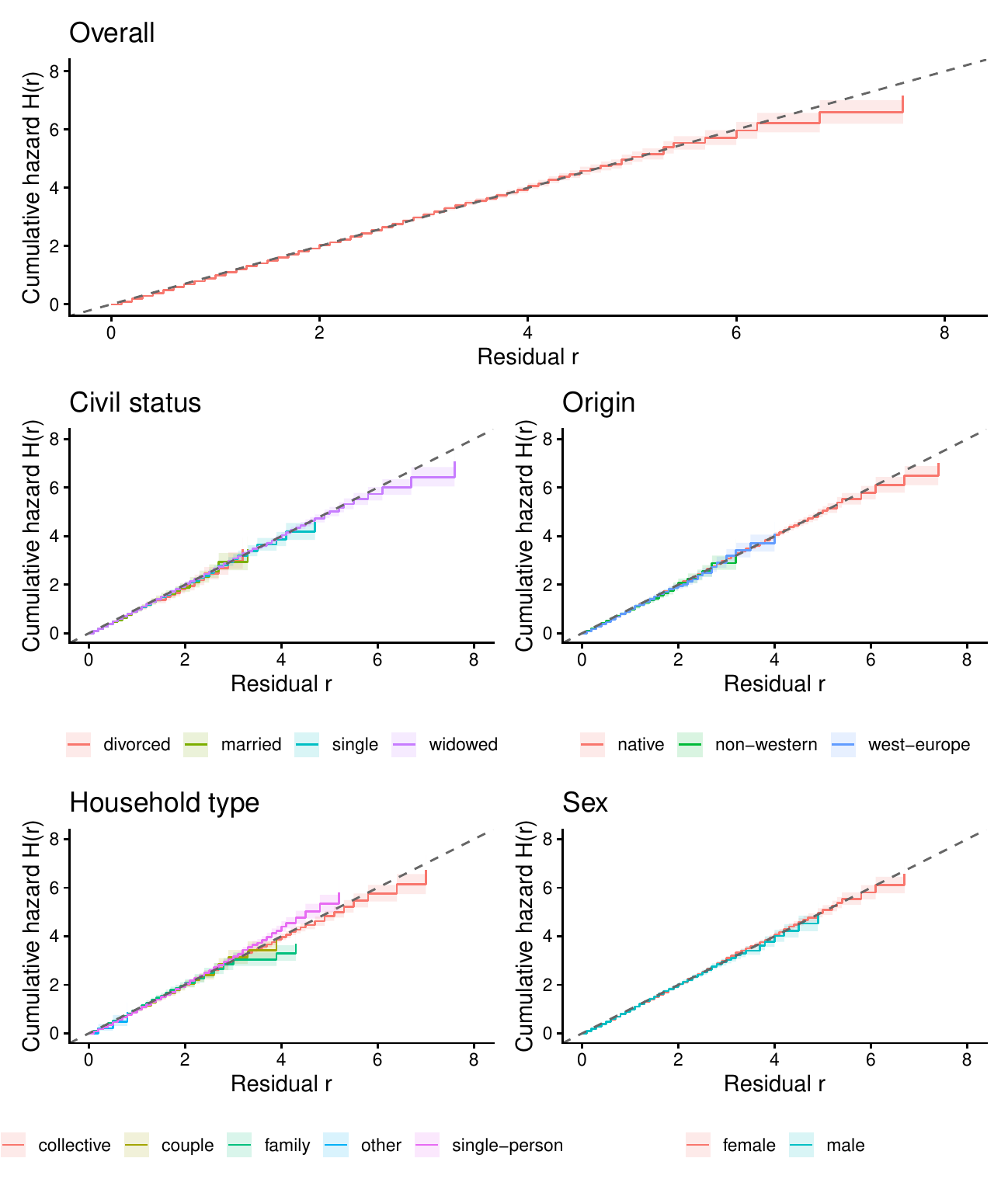}
    \caption{\added{Cumulative hazard diagnostics based on Cox--Snell residuals for the final covariate model in the Netherlands, overall and stratified by the covariates in the final specification. The dashed line indicates the 45-degree reference line implied by a correctly specified model. Shaded areas denote pointwise 95$\%$ confidence bands. Source: CBS and authors' calculations.}}
    \label{figA:diagnosticsNL}
\end{figure}

\added{Figures~\ref{figA:diagnosticsBE} and~\ref{figA:diagnosticsNL} show the resulting diagnostic plots for Belgium and the Netherlands, respectively, both overall and stratified by each covariate in the final specification. The stratified results by covariate allow us to assess whether the model also fits well within each socio-demographic subgroup, rather than only overall. For confidentiality reasons, we evaluate the cumulative hazard not at individual event times, but on a fixed grid of residual values. Adjacent grid points are merged until at least ten deaths fall between two consecutive points, and points at which fewer than ten individuals remain at risk are omitted. In all panels, the estimated cumulative hazard closely tracks the 45-degree reference line across the reported range of residual values. The pointwise 95 \% confidence bands only widen slightly in the far tail, where risk sets are small in size. We conclude that the final covariate model provides an adequate description of the data, both overall and within socio-demographic subgroups.}

\subsection{Sensitivity of parameter estimates for different threshold ages} \label{app:sensitivity}
We fit the extreme value theory framework to different threshold ages in order to assess the sensitivity of the estimated shape parameter $\xi$ as well as the estimated socio-demographic effects in the log-scale parameter $\log(\sigma_i)$. This analysis allows us to evaluate to what extent our main conclusions depend on the choice of the threshold age.

\begin{figure}[!ht]
    \centering
    \includegraphics[width=0.45\linewidth]{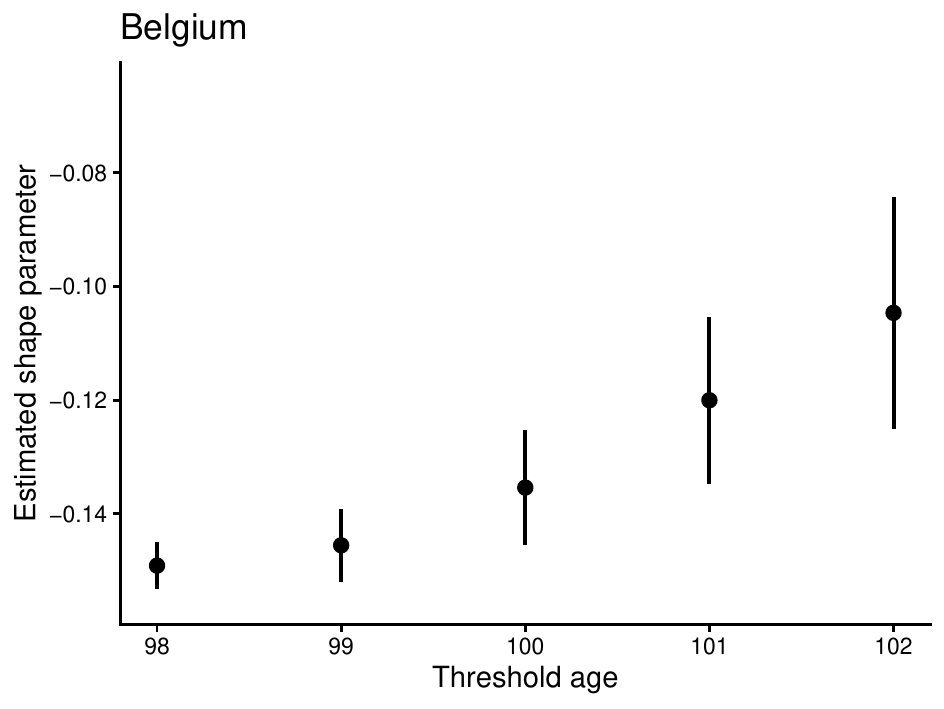}
    \includegraphics[width=0.45\linewidth]{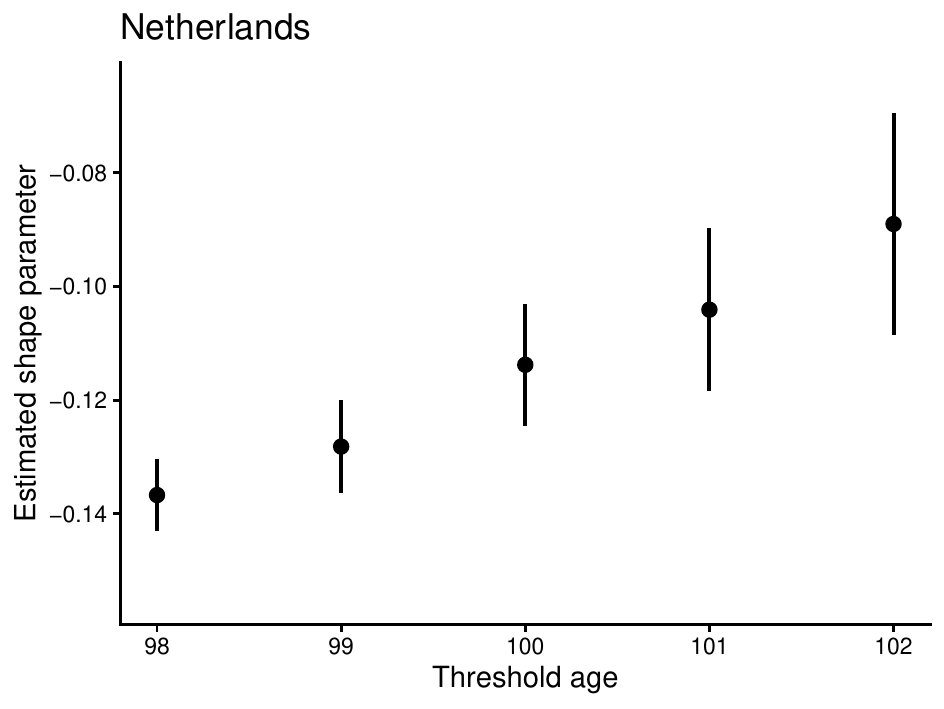}
    \caption{\added{Estimated shape parameter $\boldsymbol{\xi}$ of the Generalized Pareto distribution for different threshold ages (i.e., ages 98-102) in the Belgian (left) and Dutch (right) microdata. Point estimates are shown together with 95$\%$ confidence intervals. Source: Statbel, CBS, and authors' calculations. \label{fig:thshapesens}}}
\end{figure}

Figure~\ref{fig:thshapesens} shows the estimated shape parameters of the generalized Pareto distribution for threshold ages 98--102 in Belgium (left panel) and the Netherlands (right panel). Although there appears to be a mild increasing trend in the point estimates of the shape parameter as the threshold age increases, the width of the confidence intervals increases substantially for higher thresholds, reflecting the smaller number of exceedances. Furthermore, we observe that for all considered threshold ages the estimated shape parameter remains negative. This consistently indicates a finite upper endpoint of the distribution and is therefore consistent with the presence of a limit to the human lifespan. 

\begin{sidewaysfigure}
\centering
\includegraphics[width=\textheight]{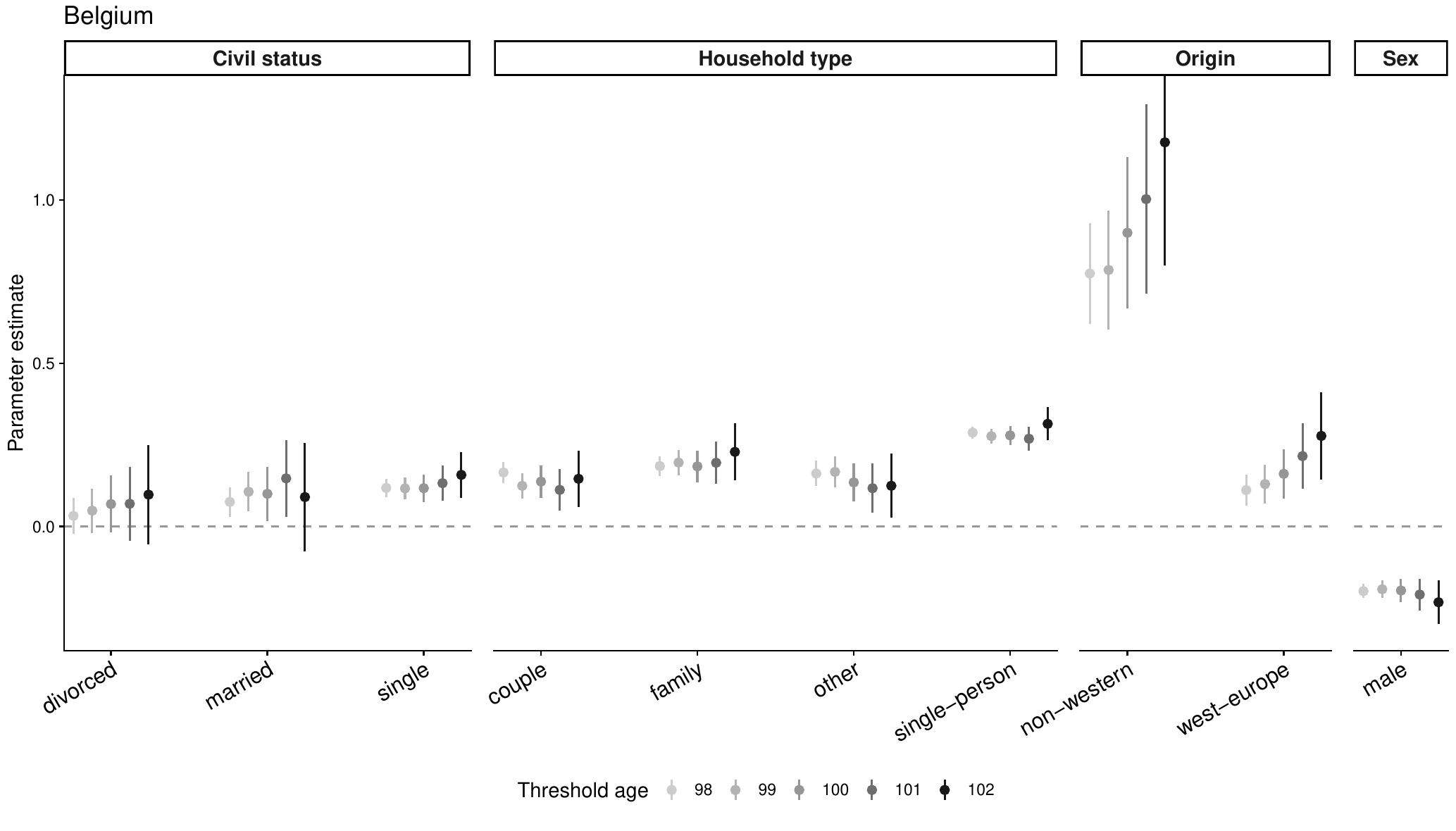}
\caption{\added{Estimated regression coefficients $\boldsymbol{\beta}_{\sigma}$ in the log-scale parameter $\log(\sigma_i) = \beta_0 + \boldsymbol{\beta}_\sigma^\top \mathbf{z}_i$ of the Generalized Pareto distribution for different threshold ages (i.e., ages 98-102) in Belgium. Point estimates are shown together with 95$\%$ confidence intervals. The point estimates of the intercept in the model for the scale parameter decrease with increasing threshold age (from 0.887 at age 98 to 0.585 at age 102). Source: Statbel and authors' calculations.\label{fig:thsensBE}}}
\end{sidewaysfigure}

\begin{sidewaysfigure}
\centering
\includegraphics[width=\textheight]{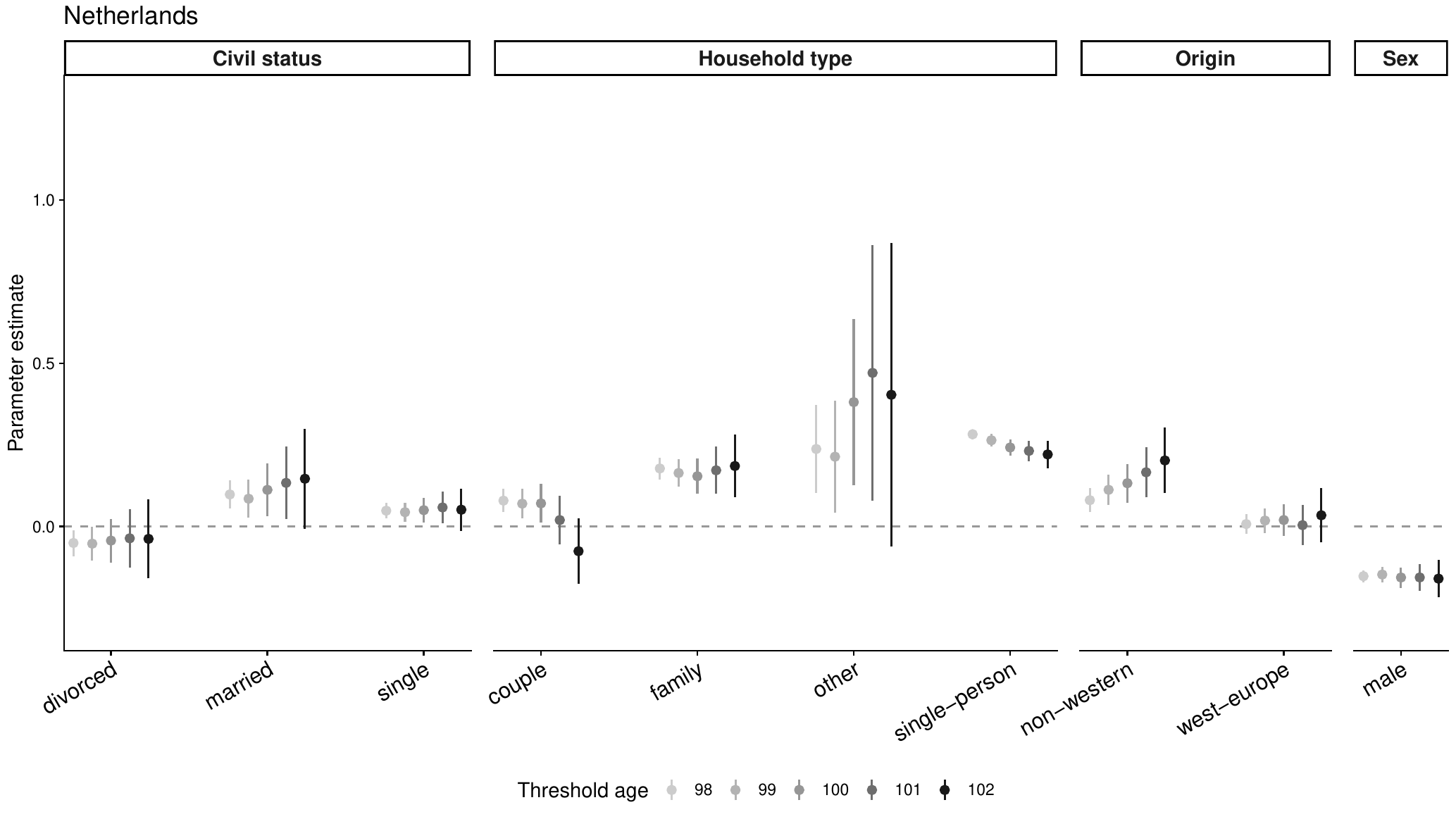}
\caption{\added{Estimated regression coefficients $\boldsymbol{\beta}_{\sigma}$ in the log-scale parameter $\log(\sigma_i) = \beta_0 + \boldsymbol{\beta}_\sigma^\top \mathbf{z}_i$ of the Generalized Pareto distribution for different threshold ages (i.e., ages 98-102) in the Netherlands. Point estimates are shown together with 95$\%$ confidence intervals. The point estimates of the intercept in the model for the scale parameter decrease with increasing threshold age (from 0.790 at age 98 to 0.509 at age 102). Source: CBS and authors' calculations.\label{fig:thsensNL}}}
\end{sidewaysfigure}

Figures~\ref{fig:thsensBE} and~\ref{fig:thsensNL} present the estimated regression coefficients in the log-scale parameter of the Generalized Pareto distribution for threshold ages 98--102 in Belgium and the Netherlands, respectively. The point estimates of the intercept in the model for the scale parameter gradually decrease with increasing threshold age, which (partly) offsets the corresponding increases in the shape parameter in terms of the implied maximum lifespan.

For the remaining covariates, the estimated coefficients display a largely stable pattern across threshold ages. The main effect of increasing the threshold age is a moderate increase in the estimation uncertainty, again due to the reduced number of exceedances. As a result, some coefficients lose statistical significance at the highest thresholds, but their point estimates remain close to those obtained at threshold age 100. Hence, the conclusions regarding socio-demographic effects remain unchanged.

Only for the covariate \texttt{org} in Belgium does there appear to be a more pronounced upward trend in the point estimates. This behavior is driven by the limited number of individuals with a non-Western European origin in the sample and by the fact that several of the most extreme ages-at-death correspond to individuals with a non-native origin. Consequently, small changes in the set of exceedances at very high thresholds have a relatively stronger impact on the estimated coefficient.

\subsection{Sensitivity to the inclusion of education in the scale parameter} \label{app:sensedu}
\added{The forward stepwise selection procedure based on the AIC excludes educational attainment from the scale parameter in the main specification. To assess whether this exclusion affects our conclusions, we refit the generalized Pareto model with education added to the retained covariates (civil status, household type, origin, and sex) at the threshold age of 100 years. Tables~\ref{tabA:sens_edu} and~\ref{tabA:sens_eduNL} compare the estimated parameters of both specifications for Belgium and the Netherlands, respectively.}
 
\added{The results are highly stable in both countries. The estimated shape parameter is essentially unchanged when we add education to the model. It equals -0.1138 without education versus -0.1140 with education in the Netherlands, and  -0.1354 versus -0.1340 in Belgium, and remains significantly different from zero in all specifications.}

\added{The estimated effects of the retained socio-demographic covariates on the scale parameter also do not change much, and no coefficient changes sign or significance status. The educational effects themselves are small. Only in Belgium, we do find that tertiary education is associated with a significantly higher scale parameter, while in the Netherlands none of the educational categories differ significantly from the primary-education reference group. In terms of model fit, adding education increases the AIC in both countries. We therefore conclude that the exclusion of education from the main specification does not affect our main results and conclusions.}
 
\begin{table}[!ht]
\centering
\adjustbox{max width = 0.68\textwidth}{
\begin{tabular}{lcc @{\hspace{0.75cm}} cc}
\toprule
& \multicolumn{2}{c}{\textbf{Without education (main)}} & \multicolumn{2}{c}{\textbf{With education (sensitivity)}}\\
 & Estimate & 95\% CI & Estimate & 95\% CI\\
\midrule
\textbf{Intercept} & 0.750$^{*}$ & (0.727, 0.774) & 0.742$^{*}$ & (0.715, 0.769)\\
\addlinespace
\multicolumn{5}{l}{\textbf{Civil status}}\\
Single & 0.118$^{*}$ & (0.076, 0.160) & 0.108$^{*}$ & (0.066, 0.151)\\
Married & 0.100$^{*}$ & (0.017, 0.184) & 0.098$^{*}$ & (0.014, 0.182)\\
Divorced & 0.069 & (-0.019, 0.157) & 0.067 & (-0.021, 0.155)\\
\addlinespace
\multicolumn{5}{l}{\textbf{Education}}\\
Secondary & --- & --- & 0.021 & (-0.010, 0.052)\\
Tertiary & --- & --- & 0.068$^{*}$ & (0.020, 0.117)\\
Unobserved & --- & --- & -0.007 & (-0.039, 0.024)\\
\addlinespace
\multicolumn{5}{l}{\textbf{Household type}}\\
Single-person & 0.279$^{*}$ & (0.251, 0.308) & 0.279$^{*}$ & (0.250, 0.307)\\
Couple & 0.138$^{*}$ & (0.088, 0.188) & 0.135$^{*}$ & (0.085, 0.186)\\
Family & 0.184$^{*}$ & (0.135, 0.233) & 0.181$^{*}$ & (0.132, 0.230)\\
Other & 0.135$^{*}$ & (0.077, 0.193) & 0.136$^{*}$ & (0.078, 0.194)\\
\addlinespace
\multicolumn{5}{l}{\textbf{Origin}}\\
West-Europe & 0.162$^{*}$ & (0.086, 0.238) & 0.164$^{*}$ & (0.088, 0.241)\\
Non-western & 0.900$^{*}$ & (0.669, 1.131) & 0.905$^{*}$ & (0.674, 1.137)\\
\addlinespace
\multicolumn{5}{l}{\textbf{Sex}}\\
Male & -0.196$^{*}$ & (-0.232, -0.160) & -0.202$^{*}$ & (-0.239, -0.166)\\
\bottomrule
\end{tabular}}
\caption{\added{Estimated parameters of the generalized Pareto model at threshold age 100 in Belgium, without education in the scale parameter (main specification) and with education (sensitivity specification), with 95$\%$ confidence intervals based on the observed Fisher information matrix. Asterisks indicate estimates for which the confidence interval excludes zero. Source: Statbel and authors' calculations.
\label{tabA:sens_edu}}}
\end{table}

\begin{table}[!ht]
\centering
\adjustbox{max width = 0.68\textwidth}{
\begin{tabular}{lcc @{\hspace{0.75cm}} cc}
\toprule
& \multicolumn{2}{c}{\textbf{Without education (main)}} & \multicolumn{2}{c}{\textbf{With education (sensitivity)}}\\
 & Estimate & 95\% CI & Estimate & 95\% CI\\
\midrule
\textbf{Intercept} & 0.650$^{*}$ & (0.629, 0.671) & 0.653$^{*}$ & (0.596, 0.711)\\
\addlinespace
\multicolumn{5}{l}{\textbf{Civil status}}\\
Single & 0.050$^{*}$ & (0.012, 0.088) & 0.049$^{*}$ & (0.010, 0.087)\\
Married & 0.113$^{*}$ & (0.032, 0.193) & 0.114$^{*}$ & (0.033, 0.194)\\
Divorced & -0.043 & (-0.110, 0.024) & -0.042 & (-0.109, 0.025)\\
\addlinespace
\multicolumn{5}{l}{\textbf{Education}}\\
Secondary & --- & --- & -0.004 & (-0.065, 0.057)\\
Tertiary & --- & --- & -0.013 & (-0.076, 0.051)\\
Unobserved & --- & --- & -0.001 & (-0.058, 0.056)\\
\addlinespace
\multicolumn{5}{l}{\textbf{Household type}}\\
Single-person & 0.242$^{*}$ & (0.218, 0.266) & 0.243$^{*}$ & (0.219, 0.267)\\
Couple & 0.071$^{*}$ & (0.012, 0.130) & 0.071$^{*}$ & (0.012, 0.131)\\
Family & 0.154$^{*}$ & (0.100, 0.208) & 0.154$^{*}$ & (0.100, 0.208)\\
Other & 0.381$^{*}$ & (0.126, 0.637) & 0.381$^{*}$ & (0.126, 0.636)\\
\addlinespace
\multicolumn{5}{l}{\textbf{Origin}}\\
West-Europe & 0.020 & (-0.028, 0.068) & 0.020 & (-0.028, 0.068)\\
Non-western & 0.133$^{*}$ & (0.074, 0.192) & 0.133$^{*}$ & (0.073, 0.192)\\
\addlinespace
\multicolumn{5}{l}{\textbf{Sex}}\\
Male & -0.156$^{*}$ & (-0.187, -0.125) & -0.155$^{*}$ & (-0.187, -0.124)\\

\bottomrule
\end{tabular}}
\caption{\added{Estimated parameters of the generalized Pareto model at threshold age 100 in the Netherlands, without education in the scale parameter (main specification) and with education (sensitivity specification), with 95$\%$ confidence intervals based on the observed Fisher information matrix. Asterisks indicate estimates for which the confidence interval excludes zero. Source: CBS and authors' calculations.
\label{tabA:sens_eduNL}}}
\end{table}

\subsection{Sensitivity to a covariate-dependent shape parameter} \label{app:sensshape}
\added{The main specification imposes a common shape parameter $\xi$ across all individuals, while we only let socio-demographic heterogeneity enter through the scale parameter $\sigma_i$. To see how strongly the covariates enter the scale, Table~\ref{tabA:sel_scale} reports the forward stepwise selection path on the scale parameter. We find that each retained covariate lowers the AIC substantially, and that educational attainment is not retained because it does not lower the AIC further for both Belgium and the Netherlands.}

\begin{table}[!ht]
\centering
\adjustbox{max width = \textwidth}{
\begin{tabular}{ll cc @{\hspace{1.5cm}} ll cc}
\toprule
\multicolumn{4}{c}{\textbf{Belgium}} & \multicolumn{4}{c}{\textbf{The Netherlands}}\\
Step & Added & AIC & $\Delta$AIC & Step & Added & AIC & $\Delta$AIC\\
\midrule
0 & Intercept only  & 33\,635.0 & ---    & 0 &  Intercept only & 39\,998.7 & ---    \\
1 & Household type  & 33\,473.4 & -161.6 & 1 & Household type  & 39\,811.7 & -187.0 \\
2 & Origin          & 33\,420.3 & -53.1  & 2 & Sex             & 39\,768.6 & -43.1  \\
3 & Sex             & 33\,370.8 & -49.5  & 3 & Origin          & 39\,761.7 & -6.9   \\
4 & Civil status    & 33\,357.6 & -13.2  & 4 & Civil status    & 39\,759.4 & -2.3   \\
\bottomrule
\end{tabular}}
\caption{\added{Forward stepwise selection path on the scale parameter at threshold age 100, with a common shape parameter. At each step the covariate yielding the largest reduction in the AIC is added, and $\Delta$AIC reflects the reduction relative to the previous step. Educational attainment is not retained, as it does not lower the AIC further. Source: Statbel, CBS, and authors' calculations. \label{tabA:sel_scale}}}
\end{table}

\added{We then assess the common-shape assumption by refitting the model while allowing the shape parameter to depend on socio-demographic covariates through a linear predictor, keeping the scale specification fixed at the retained covariates (civil status, household type, origin, and sex). We consider specifications in which each covariate enters the shape parameter individually, as well as a specification including all covariates simultaneously. Table~\ref{tabA:sens_shape} reports the AIC of each specification relative to the common-shape baseline specification.}

\begin{table}[!ht]
\centering
\adjustbox{max width = \textwidth}{
\begin{tabular}{l cc @{\hspace{0.75cm}} cc}
\toprule
& \multicolumn{2}{c}{\textbf{Belgium}} & \multicolumn{2}{c}{\textbf{The Netherlands}}\\
Shape specification & AIC & $\Delta$AIC & AIC & $\Delta$AIC\\
\midrule
Common shape (baseline)      & 33\,357.6 & --- & 39759.4 & ---\\
\addlinespace
Shape $\sim$ civil status    & 33\,363.0 & 5.3 & 39\,762.4 & 3.0\\
Shape $\sim$ education       & 33\,357.7 & 0.1 & 39\,764.2 & 4.8\\
Shape $\sim$ household type  & 33\,360.0 & 2.3 & 39\,754.1 & -5.3\\
Shape $\sim$ origin          & 33\,359.4 & 1.7 & 39\,762.0 & 2.6\\
Shape $\sim$ sex             & 33\,359.3 & 1.7 & 39\,761.3 & 1.9\\
\addlinespace
Shape $\sim$ all covariates  & 33\,366.3 & 8.6 & 39\,764.1 & 4.7\\
\bottomrule
\end{tabular}}
\caption{\added{AIC of the generalized Pareto model with covariate-dependent shape parameter at threshold age 100, relative to the common-shape baseline (main specification). The scale specification is fixed at the main-model covariates in all fits. $\Delta$AIC is the change relative to the baseline, with positive values indicating a worse fit. Source: Statbel, CBS, and authors' calculations. \label{tabA:sens_shape}}}
\end{table}
 
\added{In Belgium, none of the covariate-dependent shape specifications attains a lower AIC than the common-shape baseline. In the Netherlands, the only improvement arises when the shape parameter is allowed to depend on household type, which lowers the AIC by 5.3 points. This reduction is an order of magnitude smaller than those obtained when covariates enter the scale parameter, where household type alone lowers the AIC by 187 points (Table~\ref{tabA:sel_scale}). We therefore find no meaningful support for subgroup-specific tail behaviour, and retain a common shape parameter across socio-demographic groups in the main specification. A common shape parameter also keeps the tail behaviour comparable across the two countries and simplifies interpretation, since the scale parameter then maps directly to the estimated maximum lifespan.}

\subsection{Sensitivity to the COVID-19 period} \label{app:senscovid}
\added{Mortality at the oldest ages was strongly affected by the COVID-19 pandemic. To assess whether our results are driven by this period, we refit the final model on a sample that excludes the pandemic years (1995--2019). The pre-COVID sample contains $17\,869$ exceedances above the threshold age of 100 in Belgium compared to $22\,170$ exceedances in the full sample. For the Netherlands, the pre-COVID sample contains $22\,533$ exceedances compared to $26\,945$ exceedances in the full sample. Tables~\ref{tabA:sens_covid} and~\ref{tabA:sens_covidNL} compare the estimated parameters of both fits for Belgium and the Netherlands, respectively.}

\added{The estimates are similar across the two samples in both countries. The estimated shape parameter equals -0.1354 on the full sample and -0.1409 on the pre-COVID sample in Belgium, and -0.1138 versus -0.1123 in the Netherlands, and remains negative and significantly different from zero in all cases. The estimated socio-demographic effects in the scale parameter are very similar across both samples in Belgium and the Netherlands and the confidence intervals of the two samples overlap for all parameters. As expected, the confidence intervals are somewhat wider on the pre-COVID sample, reflecting the smaller number of exceedances. We conclude that the exclusion of the pandemic years does not affect our main conclusions.}
 
\begin{table}[!ht]
\centering
\adjustbox{max width = \textwidth}{
\begin{tabular}{lcc @{\hspace{0.75cm}} cc}
\toprule
& \multicolumn{2}{c}{\textbf{Full sample}} & \multicolumn{2}{c}{\textbf{Pre-COVID sample}}\\
 & Estimate & 95\% CI & Estimate & 95\% CI\\
\midrule
\textbf{Intercept} & 0.750$^{*}$ & (0.727, 0.774) & 0.758$^{*}$ & (0.733, 0.784)\\
\addlinespace
\multicolumn{5}{l}{\textbf{Civil status}}\\
Single & 0.118$^{*}$ & (0.076, 0.160) & 0.138$^{*}$ & (0.092, 0.184)\\
Married & 0.100$^{*}$ & (0.017, 0.184) & 0.114$^{*}$ & (0.019, 0.209)\\
Divorced & 0.069 & (-0.019, 0.157) & 0.054 & (-0.043, 0.151)\\
\addlinespace
\multicolumn{5}{l}{\textbf{Household type}}\\
Single-person & 0.279$^{*}$ & (0.251, 0.308) & 0.284$^{*}$ & (0.252, 0.315)\\
Couple & 0.138$^{*}$ & (0.088, 0.188) & 0.133$^{*}$ & (0.080, 0.187)\\
Family & 0.184$^{*}$ & (0.135, 0.233) & 0.173$^{*}$ & (0.120, 0.226)\\
Other & 0.135$^{*}$ & (0.077, 0.193) & 0.137$^{*}$ & (0.075, 0.199)\\
\addlinespace
\multicolumn{5}{l}{\textbf{Origin}}\\
West-Europe & 0.162$^{*}$ & (0.086, 0.238) & 0.203$^{*}$ & (0.116, 0.289)\\
Non-western & 0.900$^{*}$ & (0.669, 1.131) & 0.921$^{*}$ & (0.652, 1.190)\\
\addlinespace
\multicolumn{5}{l}{\textbf{Sex}}\\
Male & -0.196$^{*}$ & (-0.232, -0.160) & -0.190$^{*}$ & (-0.230, -0.150)\\
\bottomrule
\end{tabular}}
\caption{\added{Estimated parameters in the scale parameter of the generalized Pareto model at threshold age 100 in Belgium, estimated on the full sample and on the pre-COVID sample (1995-2019), with 95$\%$ confidence intervals based on the observed Fisher information matrix. Asterisks indicate estimates for which the confidence interval excludes zero. Source: Statbel and authors' calculations. \label{tabA:sens_covid}}}
\end{table}

\begin{table}[!ht]
\centering
\adjustbox{max width = \textwidth}{
\begin{tabular}{lcc @{\hspace{0.75cm}} cc}
\toprule
& \multicolumn{2}{c}{\textbf{Full sample}} & \multicolumn{2}{c}{\textbf{Pre-COVID sample}}\\
 & Estimate & 95\% CI & Estimate & 95\% CI\\
\midrule
\textbf{Intercept} & 0.650$^{*}$ & (0.629, 0.671) & 0.664$^{*}$ & (0.640, 0.687) \\
\addlinespace
\multicolumn{5}{l}{\textbf{Civil status}}\\
Single   & 0.050$^{*}$  & (0.012, 0.088)  & 0.054$^{*}$ & (0.013, 0.095)   \\
Married  & 0.113$^{*}$  & (0.032, 0.193)  & 0.105$^{*}$ & (0.014, 0.195)   \\
Divorced & -0.043 & (-0.110, 0.024) & -0.046 & (-0.121, 0.029) \\
\addlinespace
\multicolumn{5}{l}{\textbf{Household type}}\\
Single-person & 0.242$^{*}$ & (0.218, 0.266) & 0.237$^{*}$ & (0.210, 0.264)  \\
Couple        & 0.071$^{*}$ & (0.012, 0.130) & 0.059 & (-0.006, 0.123) \\
Family        & 0.154$^{*}$ & (0.100, 0.208) & 0.144$^{*}$ & (0.086, 0.202)  \\
Other         & 0.381$^{*}$ & (0.126, 0.637) & 0.306$^{*}$ & (0.027, 0.585)  \\
\addlinespace
\multicolumn{5}{l}{\textbf{Origin}}\\
West-Europe & 0.020  & (-0.028, 0.068) & 0.014 & (-0.037, 0.066) \\
Non-western & 0.133$^{*}$  & (0.074, 0.192)  & 0.156$^{*}$ & (0.088, 0.225)  \\
\addlinespace
\multicolumn{5}{l}{\textbf{Sex}}\\
Male        & -0.156$^{*}$ & (-0.187, -0.125) & -0.140$^{*}$ & (-0.175, -0.105)\\
\bottomrule
\end{tabular}}
\caption{\added{Estimated parameters in the scale parameter of the generalized Pareto model at threshold age 100 in the Netherlands, estimated on the full sample and on the pre-COVID sample (1995-2019), with 95$\%$ confidence intervals based on the observed Fisher information matrix. Asterisks indicate estimates for which the confidence interval excludes zero. Source: CBS and authors' calculations. \label{tabA:sens_covidNL}}}
\end{table}

\end{document}